\documentclass[%
superscriptaddress,
twocolumn,
 amsmath,amssymb,
prl,
]{revtex4-1}
\usepackage{appendix}
\usepackage{xcolor}
\usepackage{graphicx}
\usepackage{dcolumn}
\usepackage{bm}
\usepackage{cancel}
\usepackage{caption}
\usepackage{dsfont}
\usepackage{color} 
\newcommand{\re}{\color{black}} 
\newcommand{\jan}{\color{black}}
\newcommand{\janfin}{\color{black}}

\newcommand{\wpp}{\raisebox{.15\baselineskip}{\Large\ensuremath{\wp}}}
\newcommand{\bt}{\textbf{x}(\tau)}

\newcommand{\btt}{\tilde{\textbf{x}}(\tau)}
\newcommand{\btto}{\tilde{\textbf{x}}_0(\tau)}
\newcommand{\ud}{\mathrm{d}}
\usepackage[export]{adjustbox}
\usepackage{subcaption}
\begin{document}

\title{Thermodynamics of structure-forming systems}

\author{Jan Korbel}
 \affiliation{
 Section for the Science of Complex Systems, CeMSIIS, Medical University of Vienna, Spitalgasse 23, A-1090,
Vienna, Austria}%
\affiliation{Complexity Science Hub Vienna, Josefst\"{a}dterstrasse 39, A-1080 Vienna, Austria}

 \author{Simon David Lindner}
 \affiliation{
 Section for the Science of Complex Systems, CeMSIIS, Medical University of Vienna, Spitalgasse 23, A-1090,
Vienna, Austria}%
\affiliation{Complexity Science Hub Vienna, Josefst\"{a}dterstrasse 39, A-1080 Vienna, Austria}

\author{Rudolf Hanel}
 \affiliation{
 Section for the Science of Complex Systems, CeMSIIS, Medical University of Vienna, Spitalgasse 23, A-1090,
Vienna, Austria}%
\affiliation{Complexity Science Hub Vienna, Josefst\"{a}dterstrasse 39, A-1080 Vienna, Austria}

\author{Stefan Thurner}
\email{Correspondence to: stefan.thurner@meduniwien.ac.at}
 \affiliation{
 Section for the Science of Complex Systems, CeMSIIS, Medical University of Vienna, Spitalgasse 23, A-1090,
Vienna, Austria}%
\affiliation{Complexity Science Hub Vienna, Josefst\"{a}dterstrasse 39, A-1080 Vienna, Austria}
\affiliation{Santa Fe Institute, 1399 Hyde Park Road, Santa Fe, NM 87501, USA}
\affiliation{IIASA, Schlossplatz 1, A-2361 Laxenburg, Austria}

\date{\today} 

\begin{abstract}
{\janfin  Structure-forming systems are ubiquitous in nature, ranging from atoms building molecules to
self-assembly of colloidal amphibolic particles. The understanding of the underlying thermodynamics of such systems
remains an important problem. }
Here we derive the entropy for structure-forming systems that differs from Boltzmann-Gibbs entropy
by a term that explicitly captures `molecule' states. For large systems and low concentrations the approach is equivalent
to the grand-canonical ensemble; for small systems we find significant deviations.
We derive the detailed fluctuation theorem {\janfin and Crooks' work fluctuation theorem} for structure-forming systems.
The connection to the theory of particle self-assembly is discussed.
We apply the results to several physical systems.
We present the phase diagram for patchy particles described by the Kern-Frenkel potential.
We show that the Curie-Weiss model with molecule structures exhibits a first-order phase transition.
\end{abstract}

 \maketitle


\section{Introduction}
Ludwig Boltzmann defined entropy as the logarithm of state multiplicity.
The multiplicity of independent (but possibly interacting)
systems is typically given by multinomial factors that lead to the Boltzmann-Gibbs entropy and
the exponential growth of phase space volume as a function of the degrees of freedom.
In recent decades much attention was given to systems with long-range and co-evolving interactions that
are sometimes referred to as complex systems \cite{Thurner18}.
Many complex systems do not exhibit an exponential growth of phase space \cite{Hanel11a,Hanel11b,Hanel14,Korbel18}.
For correlated systems it typically grows sub-exponentially
\cite{Tsallis88,Rajagopal96,Kaniadakis02,Jizba04,Anteneodo04,Lutz13,Dechant15,Jizba19,Korbel20},
systems with super-exponential phase space growth were recently identified as those capable of forming
structures from its components \cite{Jensen18,Korbel18}.
A typical example of this kind are complex networks \cite{Latora17}, where complex behavior
may lead to ensemble in-equivalence \cite{Squartini15}.
The most prominent example of structure-forming systems are chemical reaction networks \cite{Berge73,Temkin96,Flamm15}.
The usual approach to chemical reactions -- where free particles may compose molecules --
is via the grand-canonical ensemble, where particle reservoirs make sure that the number of particles is conserved on average. Much  attention has been given to finite-size corrections of the chemical potential \cite{Smit89,Siepman92}  and non-equilibrium thermodynamics of small chemical networks \cite{Chandler76,Kreuzer81,Cummings84,Schmiedl07}.
However, for small closed systems, fluctuations in particle reservoirs might become non-negligible and predictions from the grand-canonical ensemble become inaccurate.
In the context of nanotechnology and colloidal physics, the theory of self-assembly \cite{Likos16} gained recent interest. Examples of self-assembly include lipid bilayers and vesicles \cite{Israelachvili77}, microtubules, and molecular motors \cite{Aranson06}, amphibolic particles \cite{Walther13}, or RNA \cite{Grabow14}. {\janfin The thermodynamics of self-assembly systems has been studied, both experimentally and theoretically, often dealing with particular systems, such as Janus particles \cite{Fantoni11}. Theoretical and computational work have explored self-assembly under non-equilibrium conditions \cite{Nguyen16,Bisker18}. A review can be found in \cite{Restrepo19}.}

Here we present a canonical approach for closed systems where particles interact and form structures.
{\janfin The main idea is to start not with a grand-canonical approach to structure forming systems
but to see within a {\em canonical description} which terms in the entropy emerge
that play the role of the chemical potential in large systems.}
A simple example for a structure-forming system, the magnetic coin model, was recently introduced in \cite{Jensen18}.
There $n$ coins are in two possible states (head and tail), and
in addition, since coins are magnetic, they can form a third state, i.e. any two coins might create a bond state.
The phase space of this model, $W(n)$,  grows super-exponentially,
$W(n)\sim n^{n/2} e^{2 \sqrt{n}} \sim e^{n \log n}$.
We first generalize this model to arbitrary cluster sizes and to an arbitrary number of states.
We then derive the entropy of the system from the corresponding
log-multiplicity and use it to compute thermodynamic quantities, such as the Helmholtz free-energy.
With respect to Boltzmann-Gibbs entropy there appears an additional term that captures the molecule states.
{\janfin
By using stochastic thermodynamics, we  obtain the appropriate second law for structure-forming systems.
and derive the detailed fluctuation theorem.
Under the assumption that external driving preserves micro-reversibility,
i.e. detailed balance of transition rates in quasi-stationary states,} we derive the non-equilibrium Crooks' fluctuation theorem for structure-forming systems. It relates the probability distribution of the stochastic work done on a non-equilibrium system to thermodynamic variables, such as the partial Helmholtz free-energy, temperature, and size of the initial and final cluster states.
Finally, we apply our results to several physical systems: We first calculate the phase diagram for the case of patchy particles described by the Kern-Frenkel potential. Second, we discuss the fully connected Ising model where molecule formation is allowed.
We show that the usual second-order transition in the fully connected Ising model changes to first-order.

\section{Results}
\subsection{Entropy of structure-forming systems}
To calculate the entropy of structure-forming systems, we first define a set of possible microstates and mesostates. Let's  consider a system of $n$ particles. Each single particle can attain states from the set $\mathcal{X}^{(1)} = \{x^{(1)}_1,\dots,x^{(1)}_{m_1}\}$. The superscript $^{(1)}$ indicates that the states correspond to a single particle state, and $m_1$ denotes the
number of these states. A typical set of states could be the spin of the particle $\{\uparrow,\downarrow\}$, or a set of energy levels. Having only single-particle states, the microstate of the system consisting of $n$ particles is a vector $(X_1,X_2,\dots,X_n)$, where $X_k \in \mathcal{X}^{(1)}$ is the state of $k$-th particle.
Let us now assume that any two particles can create a two-particle state. This two-particle state can be a molecule composed of two atoms, a cluster of two colloidal particles, etc. We call this state as a cluster. This two-particle cluster can attain states $\mathcal{X}^{(2)} =  \{x^{(2)}_1,\dots,x^{(2)}_{m_2}\}$.
A microstate of a system of $n$ particles is again a vector $(X_1,X_2,\dots,X_n)$, but now either $X_k \in \mathcal{X}^{(1)}$ or $X_k \in \mathcal{X}^{(2)} \times \mathbb{Z}_n^2$. For instance, a state of particle $k$ belonging to a two-particle cluster can be written as $X_k = x^{(2)}_1(k_1,k_2)$. The indices in the brackets tell us that the particle $k$ belongs to the cluster of size two in the state $x^{(2)}_1$ and the cluster is formed by particles $k_1$ and $k_2$ ($k_1 < k_2$). Indeed, either $k_1=k$ or $k_2=k$.

Now assume that particles can also form larger clusters up to a maximal size, $m$.
Consider $m$ as a fixed number, $m \leq n$.
Generally, clusters of size $j$ have states $\mathcal{X}^{(j)} = \{x^{(j)}_1,\dots,x^{(j)}_{m_j}\}$.
The corresponding states of the particle are always elements from sets $\mathcal{X}^{(j)} \times \mathbb{Z}_n^j$ with the restriction that if the $k$-th particle is in a state $x^{(j)}_i(k_1,\dots,k_j)$ then $k_l<k_{l+1}$, for all $l$ {\re and one} $k_l=k$. Consider an example of four particles. Particles are either in a free state, or they form a cluster of size two. A state of each particle is either $s^{(1)}$ -- a free particle, or $x^{(2)}(i,j)$ -- a cluster compound from particles $i$ and $j$. As an example, a typical microstate is $\Psi = (x^{(1)},x^{(2)}(2,3),x^{(2)}(2,3),x^{(1)})$, which means that particles $1$ and $4$ are free and particles $2$ and $3$ form a cluster.

Now consider a mesoscopic scale, where the mesostate of the system is given only by the number of clusters in each state $x^{(j)}_i$. Let us denote $n^{(j)}_i$ as the number of clusters in state $x^{(j)}_i$. The mesostate is therefore characterized by a vector $\mathds{N} = \left(n^{(j)}_i\right)$, which corresponds to a frequency (histogram) of microstates \cite{Thurner17}. The normalization condition is given by the fact that the total number of particles is $n$, i.e., $\sum_{ij} j n^{(j)}_i = n$. For example, a mesostate, $\mathds{N}_\Psi$, corresponding to a microstate $\Psi$ is $\mathds{N}_\psi = \left(n^{(1)} = 2,n^{(2)} = 1\right)$, denoting that there are two free particles and one two-particle cluster.

The Boltzmann entropy \cite{Boltzmann} of this mesostate is given by
\begin{equation}
S\left ( \mathds{N} \right) = \log W \left( \mathds{N} \right) \, ,
\end{equation}
where $W$ is the multiplicity of the mesostate, which is the number of {\em all distinct} microstates corresponding to the same mesostate.
To determine the number of {\em all distinct} microstates corresponding to a given mesostate, let us order the particles and number them from $1$ to $n$. By permutation of the particles we obtain the different possible microstates. The number of all permutations is simply $n!$. However, some permutations correspond to the same microstate and we are over-counting. In our example with one cluster and two free particles, the permutations  $(4,2,3,1)$ and $(1,3,2,4)$ correspond to the same microstate $\Psi = (x^{(1)},x^{(2)}(2,3),x^{(2)}(2,3),x^{(1)})$. However, permutation $(2,1,3,4)$ corresponds to the microstate
$\Psi' = (x^{(2)}(1,3),x^{(1)},x^{(2)}(1,3),x^{(1)})$. This microstate is a distinct microstate corresponding to the same mesostate, $\mathds{N}_\Psi \equiv \mathds{N}_{\Psi'} = \left(n^{(1)} = 2,n^{(2)} = 1\right)$.

The number of microstates giving the same mesostate can be expressed as the product of configurations with
the same state for each $x_i^{(j)}$.
Let's begin with the particles that do not form clusters.
The number of equivalent representations for one distinct state is $\left(n^{(1)}_i \right )!$, which corresponds to the number
of permutations of all particles in the same state.
For the cluster states one can think about equivalent representations
of one microstate in two steps: first permute all clusters, which gives $\left(n^{(j)}_i \right)!$ possibilities.
Then permute the particles in the cluster, which gives $j!$ possibilities for every cluster,
so that we end up with $(j!)^{n^{(j)}_i}$ combinations.

As an example, consider the case of four particles.
First, we look at free particles that attain states $x^{(1)}_1$ or $x^{(1)}_2$.  Let us consider a mesostate $\mathds{N}_1 = \left(n^{(1)}_1=2,n^{(1)}_2=2\right)$, i.e., two particles in the first state and two particles in the second.
The number of distinct microstates corresponding to the mesostate $\mathds{N}_1$ is given by $W(\mathds{N}_1) = 4!/(2!2!)=6$.
All microstates that belong to the mesostate $\mathds{N}_1$ are
\vskip 0.25cm
\begin{center}
\begin{tabular}{cc}
$(x^{(1)}_1,x^{(1)}_1,x^{(1)}_2,x^{(1)}_2) \quad$ & $\quad (x^{(1)}_1,x^{(1)}_2,x^{(1)}_1,x^{(1)}_2)$\\
$(x^{(1)}_1,x^{(1)}_2,x^{(1)}_2,x^{(1)}_1) \quad$ & $\quad (x^{(1)}_2,x^{(1)}_2,x^{(1)}_1,x^{(1)}_1)$\\
$(x^{(1)}_2,x^{(1)}_1,x^{(1)}_2,x^{(1)}_1) \quad$ & $\quad (x^{(1)}_2,x^{(1)}_1,x^{(1)}_1,x^{(1)}_2)$
\end{tabular}
\end{center}
\vskip 0.25cm
Now imagine that the four particles are either free or form two-particle clusters. The microstate of a particle is either $x^{(1)}$ or $x^{(2)}(i,j)$. Let's consider a mesostate $\mathds{N}_2 = \left(n^{(1)}=0,n^{(2)}=2\right)$, i.e., two clusters of size two. The number of distinct microstates is just $W(\mathds{N}_2)= 4!/(2! (2!)^2) =3$. The microstates corresponding to the mesostate $\mathds{N}_2$ are

\begin{center}
\begin{tabular}{c}
$(x^{(2)}(1,2),x^{(2)}(1,2),x^{(2)}(3,4),x^{(2)}(3,4))$\\
$(x^{(2)}(1,3),x^{(2)}(2,4),x^{(2)}(1,3),x^{(2)}(2,4))$\\
$(x^{(2)}(1,4),x^{(2)}(2,3),x^{(2)}(2,3),x^{(2)}(1,4))$
\end{tabular}
\end{center}

For example, a microstate $(x^{(2)}(2,1),x^{(2)}(2,1),x^{(2)}(4,3),x^{(2)}(4,3))$ is the same as the first microstate because we just relabel $1 \leftrightarrow 2$ and $3 \leftrightarrow 4$.
In summary, the multiplicity corresponding to $x^{(j)}_i$ is $(n^{(j)}_i)! (j!)^{n^{(j)}_i}$, and we can express the total multiplicity as
\begin{equation}\label{eq:mult}
W(\mathds{N}) = \frac{n!}{\prod_{ij} \left((n_i^{(j)})! (j!)^{n_i^{(j)}} \right)}\, .
\end{equation}
Using Stirling's formula $\log n! \approx n \log n - n$,
we get for the entropy
\begin{eqnarray}
S(\mathds{N}) &\approx& n \log n -  n \nonumber\\
&-& \sum_{ij} \left( n_{i}^{(j)} \log n_{i}^{(j)}  -   n_i^{(j)} + n_{i}^{(j)} \log j! \right) \, .
\end{eqnarray}
Using the normalization condition, $n = \sum_{ij} j n^{(j)}_i$,  and combining the first term with the remaining ones,
we get the entropy per particle in terms of ratios $\wp_i^{(j)} = n_i^{(j)}/n$,
\begin{eqnarray}
\mathcal{S}(\mathds{N}) = \frac{S(\{n^{(j)}_i\})}{n} = - \sum_{ij}\left[\frac{n_i^{(j)}}{n} \log \left(\frac{n_i^{(j)}}{n}\right)\right. \nonumber\\
-  \left.\frac{n_i^{(j)}}{n} \log\left(\frac{j!}{n^{j-1}}\right)  - \frac{n_i^{(j)}}{n}+\frac{j n_i^{(j)}}{n}\right]   \, .
\end{eqnarray}
Normalization is given by $\sum_{ij} j \wp_i^{(j)} = 1$. Therefore, $p_i^{(j)} = j \wp_i^{(j)}$
can be interpreted as the probability that a particle is a part of a cluster in state $x_i^{(j)}$.  On the other hand, the quantity $\wp_{i}^{(j)}$
is  the relative number of clusters. Since $\sum_{ij} \frac{j n_i^{(j)}}{n} =1$, we neglect the constant without changing the  thermodynamic relations.

In the remainder, we denote thermodynamic quantities per particle by calligraphic script and total quantities by normal script. We express the entropy per particle as
\begin{eqnarray}
\mathcal{S}(\wpp) &=& - \sum_{ij} \wp_i^{(j)} \left(\log \wp_i^{(j)}-1 \right) \nonumber\\
 &&- \sum_{ij} \wp_i^{(j)} \log\left(\frac{j!}{n^{j-1}}\right)\, ,
\end{eqnarray}
or equivalently in terms of the probability distribution, $p_i^{(j)}$, as
\begin{eqnarray}
\mathcal{S}(P) &=& - \sum_{ij} \frac{p_i^{(j)}}{j} \left(\log \frac{p_i^{(j)}}{j}-1\right)\nonumber\\
 && - \sum_{ij} \frac{p_i^{(j)}}{j} \log\left(\frac{j!}{n^{j-1}}\right)\, .
\end{eqnarray}

\paragraph*{Finite interaction range:}
Up to now we assumed an infinite range of interaction between particles,
which is unrealistic for chemical reactions, where only atoms within a short range form clusters.
A simple correction is obtained by dividing the system into a fixed number of boxes:
particles within the same box can form clusters, particles in different boxes can't.
We begin by calculating the multiplicity for two boxes.
For simplicity, assume that they both contain $n/2$ particles.
The multiplicity of a system with two boxes, $\tilde{W} \left ( n^{(j)}_i  \right)$, is given by the sum of all possible partitions
of $ n^{(j)}_i $ clusters with state $x_i^{(j)}$  into the first box (containing $ {}^1 n^{(j)}_i$ clusters)
and the second box (containing ${}^2 n^{(j)}_i$ clusters), such that $n^{(j)}_i = {}^1 n^{(j)}_i + {}^2 n^{(j)}_i$. The multiplicity is therefore
\begin{equation}
\tilde{W} \left( n^{(j)}_i\right ) = \sum_{ {}^1 n^{(j)}_i + {}^2 n^{(j)}_i = {n^{(j)}_i}} W\left({}^1 n^{(j)}_i \right) W\left({}^2 n^{(j)}_i\right )  \, ,
\end{equation}
where $W$ is the multiplicity in Eq. \eqref{eq:mult}.
The dominant contribution to the sum comes from the term, where ${}^1 n^{(j)}_i = {}^2 n^{(j)}_i = n^{(j)}_i/2$, so that
we can approximate the multiplicity by
$\tilde{W}(n^{(j)}_i) \approx W(n^{(j)}_i/2)^2$.
Similarly, for $b$ boxes we obtain  the multiplicity
\begin{equation}
    \tilde{W}(n^{(j)}_i) = W(n^{(j)}_i/b)^b =
\frac{[(n/b)!]^b}{\prod_{ij} \left([(n_i^{(j)}/b)!]^b (j!)^{n_i^{(j)}} \right)} \, .
\end{equation}
By defining the concentration of particles as $\bar c= {n}/{b}$, the entropy per particle becomes
\begin{eqnarray}
\mathcal{S}(\wpp) &=& - \sum_{ij} \wp_i^{(j)}  \left(\log \wp_i^{(j)}-1\right)\nonumber\\ &&- \sum_{ij} \wp_i^{(j)} \log\left(\frac{j!}{\bar  c^{j-1}}\right)  \, ,
\end{eqnarray}
or, respectively,
\begin{eqnarray}\label{eq:ent}
\mathcal{S}(P) &=& - \sum_{ij} \frac{p_i^{(j)}}{j} \left(\log \frac{p_i^{(j)}}{j}-1\right)\nonumber\\
 && - \sum_{ij} \frac{p_i^{(j)}}{j} \log\left(\frac{j!}{\bar  c^{j-1}}\right)\, .
\end{eqnarray}

Note that the entropy of structure-forming systems is both additive and extensive in the sense of Lieb and Yngvason
\cite{Lieb}. It is also concave, ensuring the uniqueness of the maximum entropy principle. For more details and connections to axiomatic frameworks, see Supplementary Discussion.

\subsection{Equilibrium thermodynamics of structure-forming systems}
We now focus on the equilibrium thermodynamics obtained, for example, by considering the maximum entropy principle.
Consider the internal energy
\begin{equation}
U(n_i^{(j)}) = \sum_{ij} n_i^{(j)} \epsilon_i^{(j)} = n \sum_{ij} \wp_{i}^{(j)} \epsilon_{i}^{(j)} = n \, \mathcal{U}(\wp_i^{(j)}) \, .
\end{equation}
Using Lagrange multipliers to maximize the functional
\begin{equation}
\mathcal{S}(\wpp) - \alpha\left(\sum_{ij} j \wp_i^{(j)}-1\right) - \beta \left( \sum_{ij}  \wp_i^{(j)} {\epsilon}_i^{(j)}-\mathcal{U}\right) \, ,
\end{equation}
leads to the equations,
\begin{equation}\label{eq:me}
- \log \hat{\wp}_i^{(j)} - \log \left(\frac{j!}{c^{j-1}}\right)- \alpha j - \beta \epsilon_i^{(j)} = 0 \, ,
\end{equation}
and the resulting distribution is
\begin{equation}\label{eq:maxent}
\hat{\wp}_i^{(j)} = \frac{c^{j-1}}{j!} \exp\left(-j \alpha  - \beta \epsilon_i^{(j)} \right) \, .
\end{equation}
Here we introduce the partial partition functions,
$ \mathcal{Z}_j = \frac{c^{j-1}}{j!} \sum_i e^{-\beta \epsilon_i^{(j)}}$,
and the quantity $\Lambda = e^{-\alpha}$.  $\Lambda$ is obtained from
\begin{equation}\label{eq:alpha}
\sum_{ij} j \hat{\wp}_{i}^{(j)} = \sum_{j=1}^m j\, \mathcal{Z}_j  \, \Lambda^j =1  \, ,
\end{equation}
which is a polynomial equation of order $m$ in $\Lambda$.
The connection with thermodynamics follows through Eq.~(\ref{eq:me}).
By multiplying with $\hat{\wp}_i^{(j)}$ and summing over $i,j$, we get
$\mathcal{S}(\wpp) -\sum_{ij} \hat{\wp}_i^{(j)}  - \alpha - \beta \, \mathcal{U} = 0$. 
Note that  $\sum_{ij} \hat{\wp}_{i}^{(j)} =  \sum_{ij} \hat{n}_{i}^{(j)}/n = M/n = \mathcal{M}$ is the number of clusters,
divided by the number of particles in the system.
The number of clusters per particle is
\begin{equation}\label{eq:m}
\mathcal{M} = \sum_{ij} \hat{\wp}_{i}^{(j)} = \sum_j \mathcal{Z}_j \, \Lambda^j \, .
\end{equation}
The Helmholz free-energy is thus obtained as
\begin{equation}\label{eq:free}
\mathcal{F} = \mathcal{U}-\frac{1}{\beta} \mathcal{S} = -\frac{\alpha}{\beta} - \frac{1}{\beta} \mathcal{M} \, .
\end{equation}
Finally, we can write the total partition function as
\begin{equation}
\mathcal{Z} = \exp(-\beta \mathcal{F}) = \frac{1}{\Lambda} \prod_{j=1}^m \exp(\Lambda^j \mathcal{Z}_j)\, .
\end{equation}

\begin{figure}
\captionsetup{singlelinecheck = false, format= hang, justification=raggedright, labelsep=space}
\includegraphics[width=\linewidth]{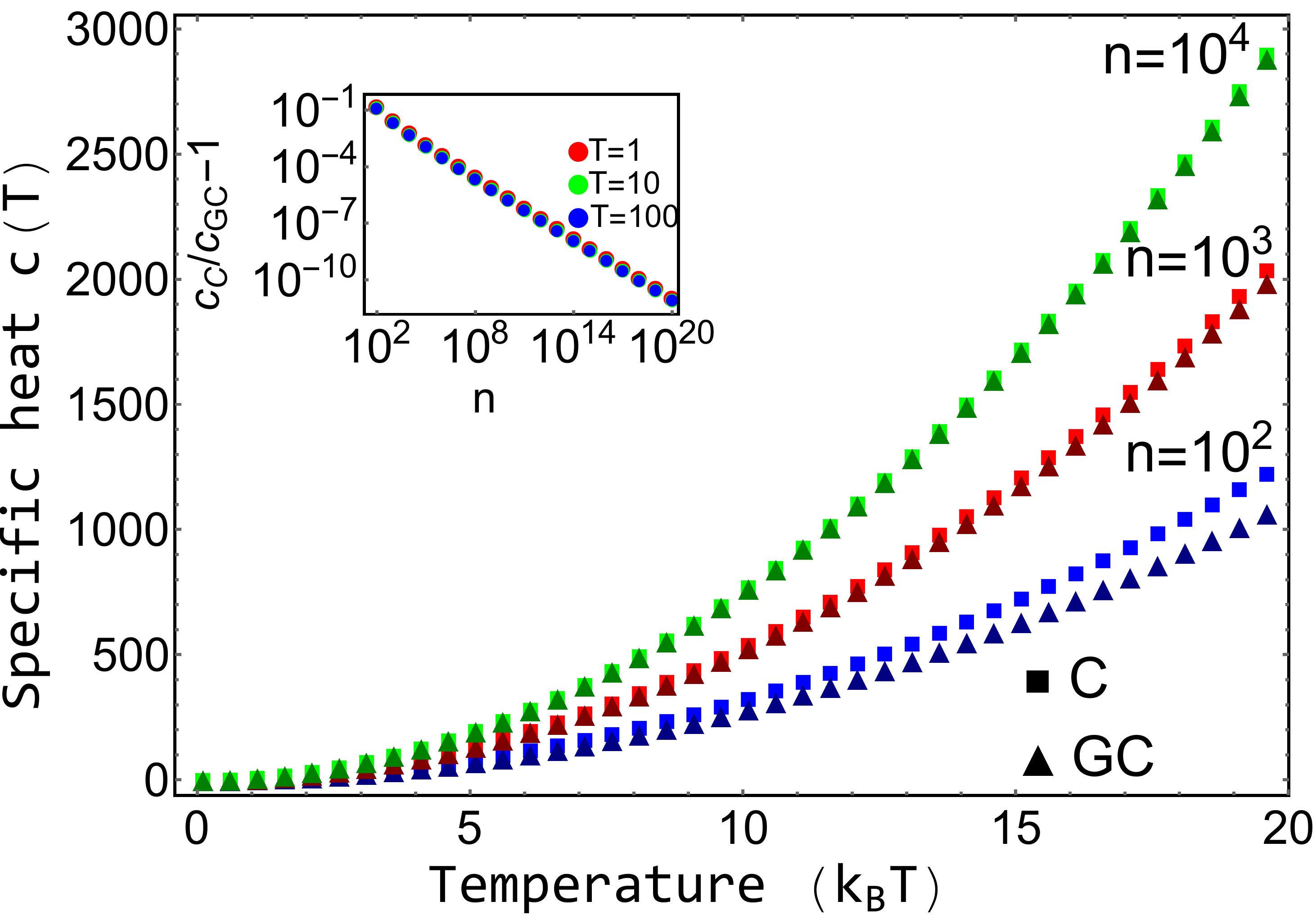}
\caption{Specific heat, $c(T)$, for the reaction $2X \rightleftharpoons X_2$ for the presented canonical approach
({\janfin C}, squares) with an exact number of particles
in comparison to the grand-canonical ensemble ({\janfin GC}, triangles).
{\janfin $n$ denotes the number of particles.}
For small systems the difference of the approaches becomes apparent.
The inset shows the ratio of the specific heat calculated from the  exact approach to the one obtained from the grand-canonical ensemble,
 $c_{C}/ {c}_{GC}-1$. For large $n$ the quantity decays to zero for any temperature.}
    \label{fig:free}
\end{figure}

\paragraph*{Comparison with the grand-canonical ensemble:}
To compare the presented exact approach with the grand-canonical ensemble,
consider the simple chemical reaction, $2 X \rightleftharpoons X_2$. Without loss of generality, assume
that free particles carry some energy, $\epsilon$.  {\janfin We calculate the Helmholtz free-energy for both approaches in the
Supplementary Information}.
In Fig. \ref{fig:free}, we show the corresponding specific heat, $c(T) = - T \frac{\partial^2 \mathcal{F}}{\partial T^2}$.
For large systems the usual grand-canonical ensemble approach and
the exact calculation with a strictly conserved number of particles converge.
For small systems, however, there appear notable differences.
This is  visible in Fig. \ref{fig:free}, where only for large $n$ and low concentrations, $\bar c$, the specific heat for the exact approach (squares) and the grand-canonical ensemble (triangles) become identical.
The inset shows the ratio of the specific heat, $c_{C}/c_{GC}-1$, vanishing for large $n$.
For large systems the exact approach and the the grand-canonical ensemble are equivalent.

\subsection{Relation to the theory of self-assembly}
In many applications, the number of energetic configurations for each cluster-size is so large
that one is only interested in the distribution of cluster-sizes. For this case it is possible to formulate an effective theory considering contributions from all configurations that is known as the theory of self-assembly. For an overview, see \cite{Likos16}.

To compute the free-energy in terms of the cluster-size distribution,
we define the latter as
\begin{equation}\label{eq:cl}
\hat{\wp}^{(j)} = \sum_i \hat{\wp}_i^{(j)} = \Lambda^j \mathcal{Z}_j \, .
\end{equation}
This is the distribution obtained from {\re a free-energy of the ideal gas} of clusters, as discussed in \cite{Fantoni11} for the case of Janus particles and in \cite{Vissersa14} for the more general case of one-patch colloids. The entropy of the relative cluster-size can be introduced as
\begin{equation}
\mathcal{S}_c(\wpp) = - \sum_{j=1}^m \wp^{(j)}  \left(\log \wp^{(j)}-1\right)\, .
\end{equation}
By introducing the partial free-energy as
\begin{equation}\label{eq:freeen}
\Phi_j= -\frac{1}{\beta} \log \mathcal{Z}_j\, ,
\end{equation}
the 
energy constraint takes the form of the expected free-energy, averaged over cluster-size,
$ \Phi =  \sum_{j=1}^m \wp^{(j)} \Phi_j$.
The cluster-size distribution is obtained by maximization of the functional
\begin{equation}
\mathcal{S}_c(\wpp) -  \alpha_c \left(\sum_{j=1}^m j \wp^{(j)} -1 \right) - \beta \left(\sum_{j=1}^m \wp^{(j)}\Phi_j  - \Phi \right)\, .
\end{equation}
It is clear that Eq. \eqref{eq:cl} is the solution of the maximization. The free-energy can be now expressed as
\begin{equation}
\mathcal{F}_c = \Phi - 1/\beta S_c = - \frac{\alpha_c}{\beta} - \frac{\mathcal{M}}{\beta} \, ,
\end{equation}
which has the same structure as when calculated in terms of $\wp_i^{(j)}$.

\subsection{Examples for thermodynamics of structure-forming systems}
We now {\janfin apply the results obtained in the previous section} to several examples of structure-forming systems.
{\janfin We particularly focus on how the presence of mescoscopic structures of clustered states leads to the macroscopic physical properties.
In the presence of structure-formation there exists a phase transition between a free particle fluid phase and a condensed phase, containing clusters of particles. This phase transition is demonstrated in two examples.}

The first example on soft-matter self-assembly describes the process of condensation of one-patch colloidal amphibolic particles.
This condensation is relevant in applications in nanomaterials and biophysics.
The second example covers the phase transition of the Curie-Weiss spin model for the situation where particles form molecules.
In the Supplementary Information we discuss the additional examples of a magnetic gas and a size-dependent chemical potential.

\begin{figure}[t]
\captionsetup{singlelinecheck = false, format= hang, justification=raggedright, labelsep=space}
    \includegraphics[width= \linewidth]{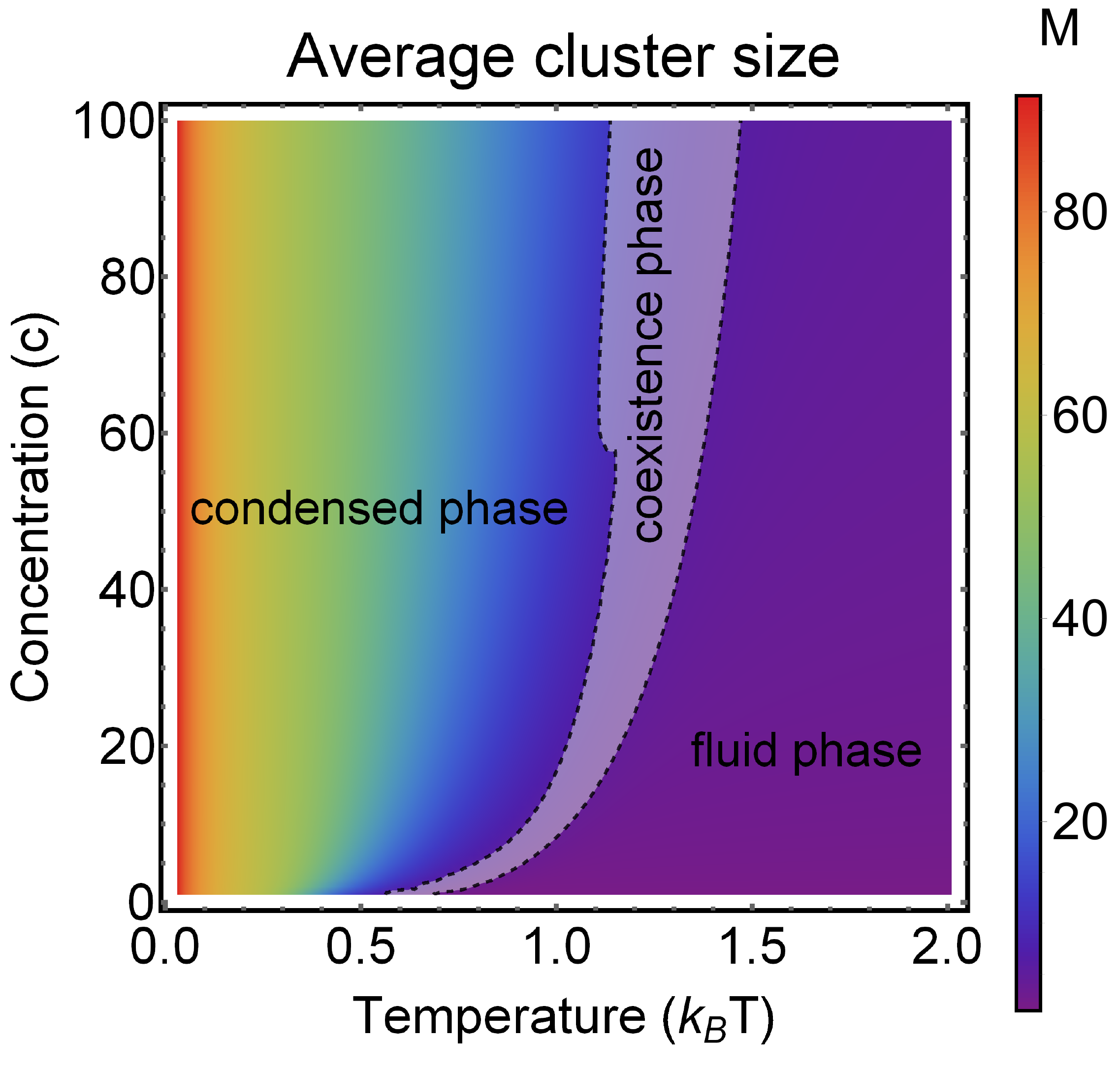}
    \caption{Phase diagram for the self-assembly of patchy particles for $n=100$ particles.
    The average cluster size ($M$) as a function of temperature ($T$) and concentration ($c$) is seen.
    {\janfin The cluster-size is given by the color and ranges from $M=0$ (purple) to $M=100$ (red). }
    We observe three phases: The liquid and condensed phase are divided by a coexistence phase (gray area).
    Coexistence is characterized by a bimodal distribution that can be detected with a shift in the bimodality coefficient.}
    \label{cluster}
\end{figure}

\paragraph*{Kern-Frenkel model of patchy particles:}
Recently, the theory of soft-matter self-assembly has successfully predicted the creation of various structures of colloidal particles, including clusters of Janus particles \cite{Fantoni11}, polymerization of colloids \cite{Vissersa14}, and the crystallization of multi-patch colloidal particles \cite{Preisler14}. Kern and Frenkel \cite{Kern03} introduced a simple model to describe the self-assembly of amphibolic particles with two-particle interactions. $\textbf{r}_{ij}$ denotes a unit vector connecting the centers of particles $i$ and $j$, $r_{ij}$ is the corresponding distance, and $\textbf{n}_i$ and $\textbf{n}_j$ are unit vectors encoding the directions of patchy spheres. The Kern-Frenkel potential was defined as
\begin{equation}
U^{KF}_{ij} = u(r_{ij}) \Omega(\textbf{r}_{ij},\textbf{n}_i,\textbf{n}_j) \, ,
\end{equation}
where
\begin{equation*}
u(r_{ij}) = \left\{ \begin{array}{rl} \infty, & r_{ij} \leq \sigma \\  - \epsilon, & \sigma < r_{ij} < \sigma + \Delta \\ 0, & r_{ij} > \sigma + \Delta. \end{array} \right.
\end{equation*}
and
\begin{equation*}
\Omega(\textbf{r}_{ij},\textbf{n}_i,\textbf{n}_j) =\left\{
                                                     \begin{array}{ll}
                                                       -1 & \hbox{if}  \left\{
                                                                         \begin{array}{ll}
                                                                           \textbf{r}_{ij} \cdot \textbf{n}_i > \cos \theta & \hbox{and} \\
                                                                           \textbf{r}_{ij} \cdot \textbf{n}_j > \cos \theta &
                                                                         \end{array}
                                                                       \right.
\\
                                                      \ \ 0, & \hbox{otherwise.}
                                                     \end{array}
                                                   \right.
\end{equation*}
The characteristic quantity, $\chi = \sin^2(\theta/2)$, is the particle coverage.
In the theory of self-assembly the cluster-size distribution is determined by the partial partition functions \eqref{eq:cl}.
Due to the enormous number of possible configurations it is impossible to calculate $\mathcal{Z}_j$ analytically and
simulation methods were introduced, including a grand-canonical Monte Carlo method, and Successive Umbrella Sampling; for a review, see \cite{Rovigatti18}. Instead of calculating the exact value of $\mathcal{Z}_j$, we use a stylized model based on \cite{Fantoni11}. There  the partial partition function is parameterized as
$\frac{\log \mathcal{Z}_j}{j \epsilon} = b \tanh (aj)$,
where $b<0$ and $a>0$ are the model parameters. While for small cluster-sizes, the free-energy per particle decreases linearly with the size, for larger clusters, it saturates at $b$. To calculate the average cluster-size Eq. \eqref{eq:m} one has to solve the equation for $\Lambda$, \eqref{eq:alpha}. In Fig. \ref{cluster}, we show the phase diagram of the patchy particles for $b = -3$ and $a=25$ and $n=100$. The average number of clusters, $M$, plays the role of the order parameter. In the phase diagram one can clearly distinguish three phases. At high temperature, we observe the liquid phase, where most particles are not bound to others.
At low temperatures we have a condensed phase with macroscopic clusters. The two phases are separated by a coexistence phase, where both, large clusters and unbounded particles are present. The coexistence phase (gray region) is characterized by a bimodal distribution that can be recognized by calculating the bimodality coefficient \cite{bimodality}. Results presented in Fig. \ref{cluster} qualitatively correspond to results obatined in \cite{Fantoni11} for the case of Janus particles with $\xi=0.5$.

\begin{figure}[t]
\captionsetup{singlelinecheck = false, format= hang, justification=raggedright, labelsep=space}
    \includegraphics[width= 0.9\linewidth]{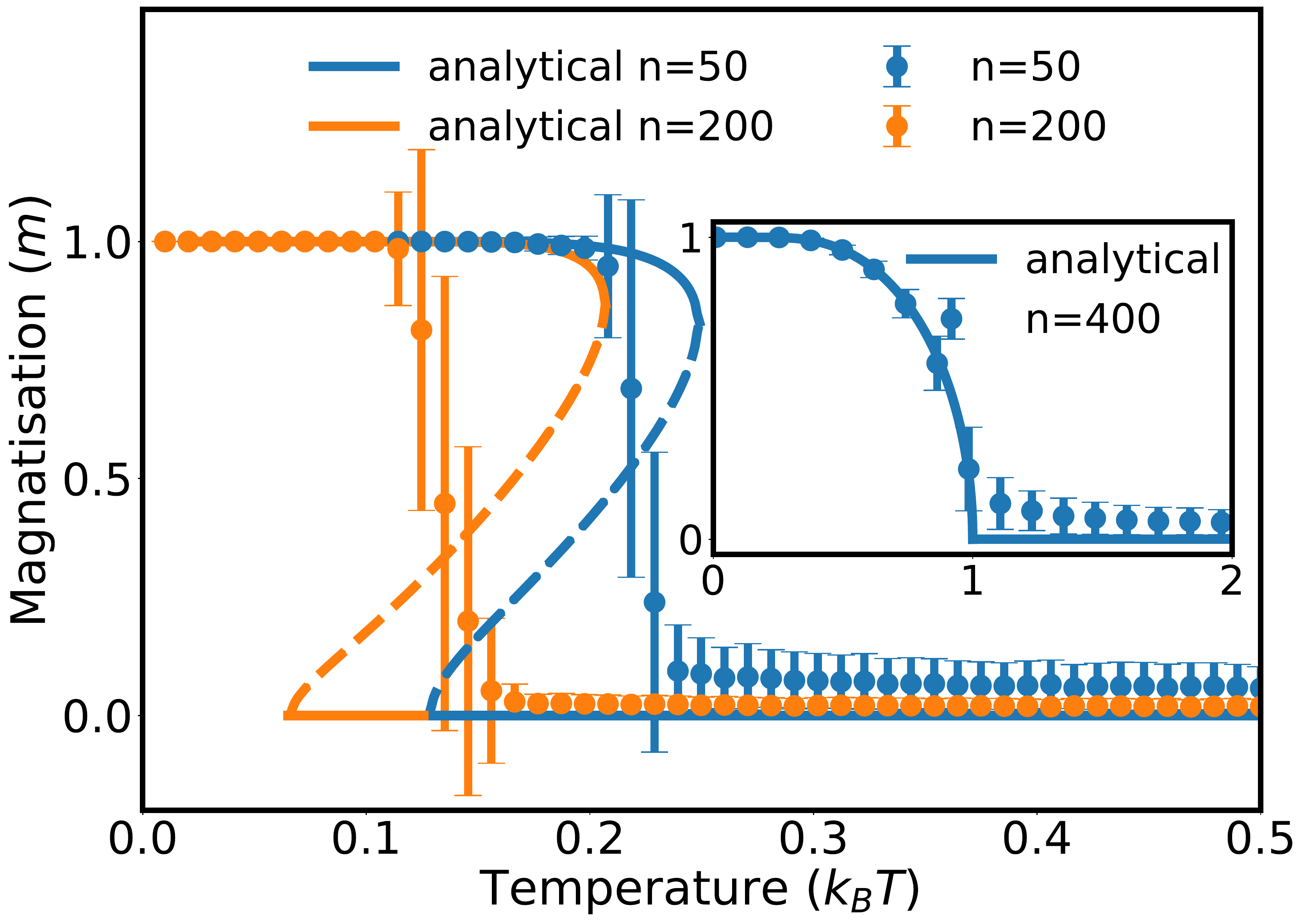}
    \caption{Magnetization of the fully connected Ising model with molecule states for
    $n=50$ and $n=200$ particles, {\janfin for a spin-spin coupling constant, $J=1$}.
    Results of the mean-field approximation (solid lines) are in good agreement with Monte-Carlo simulations (symbols).
    {\jan Errorbars show the standard deviation of the average value obtained from 1000 independent runs of the simulations
    (see Supplementary Information for more details).}
    The inset shows the well-known result for the fully connected Ising model without molecule states.
    Without molecule formation we observe the usual second-order transition.
    With molecules, the critical temperature decreases with the number of particles and the phase transition becomes first-order.}
    \label{mc_sim}
\end{figure}

\paragraph*{Curie-Weiss model with molecule formation:}
To discuss an example of a spin system with molecule states, consider the fully connected Ising model
\cite{Griffiths66,Botet82,Gulbahce04,Romano14} with a Hamiltonian that allows for possible molecule states
\begin{equation}
H(\sigma_i) = -\frac{J}{n-1} \sum_{i\neq j, \, \, \text{free}} \sigma_i \sigma_j - h \sum_{j, \, \, \text{free}} \sigma_j \, .
\end{equation}
Molecule states neither feel the spin-spin interaction nor the external magnetic field, $h$.
Therefore, the sum only extends over free particles.
In a mean-field approximation we use the magnetization, \hbox{$m = \frac{1}{n-1} \sum_{i \neq j} \sigma_i$},
and express the Hamiltonian as \mbox{$H^{MF}(\sigma_i) = -(Jm+h) \sum_{j,\text{free}}  \sigma_j$}.
The self-consistency equation $m =  -\frac{\partial F}{\partial h}|_{h=0}$ leads to an equation for $m$
that is calculated numerically (Supplementary Information) and that is shown in Fig. \ref{mc_sim}.
Contrary to the mean-field approximation of the usual fully connected Ising model (without molecule states),
the phase transition is no longer second-order but becomes first-order.
There exists a bifurcation where solutions for $m=0$ and $m>0$ are stable.
The second-order transition is recovered for small systems, $n \rightarrow 0$.
The critical temperature is shifted towards zero for increasing  $n$.
We performed Monte-Carlo simulations to check the result of  the mean-field approximation; see Supplementary Information.

\subsection{Stochastic thermodynamics of structure-forming systems}
Consider an arbitrary non-equilibrium state given by  ${\wp}_i^{(j)} \equiv {\wp}_i^{(j)}(t)$, and
and imagine that the evolution of the probability distribution is defined by a first-order Markovian
linear master equation, as is usually assumed in stochastic thermodynamics \cite{Seifert08, Esposito10}
\begin{equation} \label{eq:mem}
\dot{\wp}_i^{(j)} = \sum_{kl} w_{ik}^{jl} \wp_k^{(l)} = \sum_{kl} \left(w_{ik}^{jl} \wp_k^{(l)} - w_{ki}^{lj} \wp_i^{(j)}\right)  \, .
\end{equation}
$w_{ik}^{jl}$ are the transition rates.
Note that {\janfin probability} normalization leads to $\sum_{ij} j \dot{\wp}_i^{(j)}  = 0$.
Given that detailed balance holds,
$ w_{ik}^{jl} \hat{\wp}_k^{(l)} = w_{ki}^{lj} \hat{\wp}_i^{(j)}$,
the underlying stationary distribution, {\janfin obtained from $\dot{\wp}_i^{(j)}=0$, coincides with the equilibrium distribution \eqref{eq:maxent}.}
From this we get
\begin{eqnarray}
 \frac{w_{ik}^{jl}}{w_{ki}^{lj}} &=& \frac{j!}{l!}c^{l-j} \exp\left[\alpha(l-j)+\beta \left (\epsilon_{k }^{(l)}-\epsilon_{i}^{(j)} \right )\right] \, .
\end{eqnarray}
The time derivative of the entropy per particle is
\begin{equation}
\frac{\mathrm{d} \mathcal{S}}{\mathrm{d}t} = - \sum_{ij} \dot{\wp}_{i}^{(j)} \log \wp_{i}^{(j)} - \sum_{ij} \dot{\wp}_{i}^{(j)} \log \left(\frac{j!}{c^{j-1}}\right) \, .
\end{equation}
Using the master equation \eqref{eq:mem} and some straightforward calculations,
we end up with the usual second law of thermodynamics,
\begin{equation}
\frac{\mathrm{d} \mathcal{S}}{ \mathrm{d} t}
 =   \dot{\mathcal{S}}_i + \beta \dot{\mathcal{Q}}  \, ,
 \end{equation}
where $\dot{\mathcal{Q}}$ is the heat flow per particle and $\dot{\mathcal{S}}_i$ is the non-negative entropy production per particle, see Supplementary Information.

Let us now consider a stochastic trajectory, $\bt = (i(\tau),j(\tau))$, denoting that at time $\tau$,
the particle is in state $x_{i(\tau)}^{(j(\tau))}$.
We introduce the time-dependent protocol, $l(\tau)$, that controls the energy spectrum of the system.
The stochastic energy for  trajectory $\bt$ and protocol $l(\tau)$ can be expressed as $\epsilon(\tau) \equiv \epsilon_{i(\tau)}^{(j(\tau))}(l(\tau))$.  We assume micro-reversibility from which follows that detailed balance is valid even when the energy spectrum is time-dependent (due to protocol $l(\tau)$). We define the stochastic entropy as
\begin{equation}
s(\bt) = -  \left(\log \wp_{i(\tau)}^{(j(\tau))}(\tau)- 1\right) -  \log \left(\frac{j(\tau)!}{c^{j(\tau)-1}}\right)\, .
\end{equation}
{We show that $\dot{s}(\bt) = \dot{s}_i(\bt) +  \dot{s}_e(\bt)$, where $\dot{s}_i$ is the stochastic entropy production rate and $\dot{s}_e$ is the entropy flow equal to ${\dot{q}}/{T}$, where $\dot{q}$ is the heat flow in the Supplementary Information.

The time-reversed trajectory is $\btt = (i(T-\tau),j(T-\tau))$, and the time reversed protocol, $\tilde{l}(\tau) = l(T-\tau)$.
The log-ratio of the probability, $\mathcal{P}$, of a forward trajectory and the probability, $\tilde{\mathcal{P}}$,
of the time-reversed trajectory under the time-reversed protocol is equal to $\Delta \sigma =  \Delta s_i + \log \frac{j_0}{\widetilde{j}_0}$,
where $j_0 = j(\tau=0)$ and $\widetilde{j}_0 = \widetilde{j}(\tau=0)$, see Supplementary Information.
Hence,
$\log \frac{\mathcal{P}(\bt)}{\tilde{P}(\btt)} = \Delta \sigma$, which leads to the fluctuation theorem \cite{Esposito10a}
\begin{equation}\label{eq:ft}
\log \frac{P(\Delta \sigma)}{\tilde{P}(-\Delta \sigma)} = \Delta \sigma \, .
\end{equation}
Assuming that the initial state is an equilibrium state, introducing the stochastic free-energy, $f(\tau) = \epsilon(\tau)- T s(\tau)$, and combining the first and the second law of thermodynamics, we get
$ \Delta s_i = \beta (w - \Delta f)$. The stochastic free-energy of an equilibrium state is
$f(\hat{\wp}_i^{(j)}) = - j \frac{\alpha}{\beta} - \frac{1}{\beta}$, see Supplementary Information.

If we start in an equilibrium distribution with $j(\tau=0)=j_0$ and the reverse experiment also starts in an equilibrium distribution with $\tilde{j}(\tau=0) = \tilde{j}_0$, by plugging this into Eq. \eqref{eq:ft} and a simple manipulation we have
\begin{equation}
\frac{\mathcal{P}(\bt|j_0)}{\tilde{\mathcal{P}}(\btt|\tilde{j}_0)} = \exp\left(\beta w - \beta \left[\Phi_{\tilde{j}_0}(\tilde{l}(0)) - \Phi_{j_0}(l(0))\right]\right) \, ,
\end{equation}
where $\Phi_j$ is the partial free-energy \eqref{eq:freeen}. Finally, by a straightforward calculation we obtain
Crooks' fluctuation theorem \cite{Crooks99,Esposito10a},
\begin{equation}
\frac{P(w|j_0)}{\tilde{P}(-w|\tilde{j}_0)} = \exp(\beta (w -  \Delta \Phi_j))  \,
\end{equation}
where  $\Delta \Phi_j = \Phi_{\tilde{j}_0}(\tilde{l}(0)) - \Phi_{j_0}(l(0))$. For technical details, see Supplementary Information.

\section{Discussion} 
We presented a straight forward way to establish the thermodynamics of structure-forming systems
(e.g. molecules made from atoms or clusters of colloidal particles)
based on the {\em canonical ensemble} with a {\em modified entropy} that is obtained by the
proper counting of the system's configurations.
The approach is an alternative to the grand-canonical ensemble that yields identical results for large systems.
However, there are significant deviations that might
have important consequences for small systems, where the interaction range becomes comparable with system-size.  Note that our results are valid for large systems (in the thermodynamic limit) as well as small systems at nano-scales.
We showed that fundamental relations such as the second law of thermodynamics and fluctuation theorems remain valid for structure-forming systems.
{\re In addition, we demonstrated that the choice of a proper entropic functional has profound physical consequences. It determines, for example, the order of phase-transitions in spin models.}

We mention that we follow a similar reasoning as has been used in the case of Shannon's entropy: Originally, Shannon entropy was derived by Gibbs in the thermodynamic limit using a frequentist approach to statistics (probability is given by a large number of repetitions). However, once the formula for entropy had been derived, its validity was extended beyond the thermodynamic limit, which corresponds to the Bayesian approach. It has been shown, e.g., by methods of stochastic thermodynamics, that the formula for the Shannon entropy and the laws of thermodynamics remain valid for systems of arbitrary size (with the exception of systems with quantum corrections) and arbitrarily far from equilibrium \cite{Seifert08}. In this paper, we follow the same type of reasoning for the case of structure-forming systems.

Typical examples where our results apply are chemical reactions at small scales, the  self-assembly of colloidal particles,
active matter, and nano-particles. The presented results might also be of direct use for chemical nano-motors \cite{Kagan09}
and  non-equilibrium self-assembly \cite{Restrepo19}.
A natural question is how the framework can be extended
to the well-known statistical physics of chemical reactions \cite{Chandler76,Kreuzer81,Cummings84,Schmiedl07}
where systems are composed of more than one type of atom.

\section*{Acknowledgements}
We acknowledge support from the Austrian Science fund Projects I 3073 and P 29252
and the Austrian Research Promotion agency FFG under Project 857136.
We thank Tuan Pham for helpful discussions.

{\janfin
\section*{Data availability}
All relevant data are available at: https://github.com/complexity-science-hub/Thermodynamics-of-structure-forming-systems.

\section*{Author contributions}
J.K., R.H. and S.T. conceptualized the work, S.D.L. performed the computational work, all authors contributed to analytic calculations and wrote the paper.

\section*{Competing interests}
The authors declare no competing interests.
}

\clearpage

\onecolumngrid
\appendix

\section{Supplementary Information}

This Supplementary Information to the paper {\em Thermodynamics of {structure-forming} systems} contains
additional information, mainly on details of analytical and numerical computations. It also contains more examples that we mention in the main text.

\section{Supplementary Methods}

\subsection{Equivalence of the exact calculation of the sample space with the grand-canonical ensemble in the thermodynamic limit}

Here we show the equivalence of the presented approach with the grand-canonical ensemble in
the thermodynamic limit and the limit of low concentrations.
Let us consider a chemical reaction, $2 X \rightleftharpoons X_2$, with $n$ particles.
Let us denote the number of particles $X$ as $n_X$ and the number of molecules $X_2$ as $n_{X_2}$.
Without loss of generality, let us consider that free particles have an energy $\epsilon$ and molecules have zero energy.

\paragraph{Exact calculation:} Let us start with entropy
\begin{equation}
S({\wp}_X,{\wp}_{X_2}) = -  {\wp}_X \log { \wp}_X -  { \wp}_{X_2} (\log { \wp}_{X_2} +1) - { \wp}_{X_2} \log\left(\frac{2}{c}\right)
\end{equation}
normalization constraint, ${ \wp}_X+2 { \wp}_{X_2} = 1 $, and the energy constraint, $\epsilon { \wp}_X = \mathcal{U}$.

From this we obtain
\begin{eqnarray}
{\wp}_X &=& \exp(-(\alpha)-\beta \epsilon)\, , \\
{\wp}_{X_2} &=&  \frac{c}{2} \exp(-2(\alpha))\, .
\end{eqnarray}
The Lagrange multiplier $\alpha$ can be calculated from the normalization constraint
\begin{equation}\label{eq:alphas}
 \exp(-(\alpha)-\beta \epsilon) + 2  \cdot  \frac{c}{2} \exp(-2(\alpha)) = 1\, .
\end{equation}
We obtain two solutions of the quadratic equation, of which only one has a physical meaning, i.e.,
\begin{equation}
\alpha =\log \left( \frac{2c e^{-1+\beta \epsilon}}{-1+\sqrt{1+4c e^{2 \beta \epsilon}}} \right) \, .
\end{equation}
Helmholtz free-energy can be obtained as
\begin{equation}
F = - \frac{\alpha n}{\beta} - \frac{n({   \wp}_X+{\wp}_{X_2})}{\beta}  =  \frac{n}{\beta} \log \left( \frac{-1+\sqrt{1+4c e^{2 \beta \epsilon}}}{2c e^{\beta \epsilon-1}} \right
  ) - \frac{n}{\beta} \frac{e^{-2 \beta  \epsilon } \left(2 c e^{2 \beta  \epsilon }+\sqrt{4 c e^{2 \beta  \epsilon }+1}\right)}{4 c}
\end{equation}

\paragraph{Grand-canonical ensemble:} Let's now compare the exact result with the usual
approach using the grand-canonical ensemble.
The partition function of the grand-canonical ensemble can be expressed as
\begin{equation}
\mathbb{Z} = \sum_{n_X,n_{X_2}=0}^\infty \frac{1}{n_X!}\exp(-\beta(\epsilon-\mu_X)n_X)
 			\frac{1}{n_{X_2}!}\exp(\beta \mu_{X_2} n_{X_2})
 = \exp \left(e^{\beta  \mu_{X_2} }+e^{-\beta  (\epsilon -\mu_X )}\right) \, ,
 \end{equation}
where $\mu_X$ and $\mu_{X_2}$ are the chemical potentials.
From the Gibbs-Duhem relation, we get that $\mu_{X_2} = 2 \mu_X$.
We denote the chemical potential by $\mu$. The average number of particles can be calculated as
\begin{equation}
\langle n \rangle = \frac{\partial \log \mathbb{Z}}{\beta \partial \mu} = 2 e^{2 \beta  \mu }+e^{\beta  (\mu -\epsilon )}\, .
\end{equation}
This relation serves as an equation for $\mu$, which has the same form as
Eq. \eqref{eq:alphas}, and the solution can be found as
\begin{equation}
\mu = \frac{\log \left(\frac{-1+\sqrt{1+8 \langle n \rangle e^{2 \beta  \epsilon
   }}}{4 e^{\beta  \epsilon}}  \right)}{\beta }\, .
\end{equation}
Helmholtz free-energy can be expressed from the grand-potential
$\Omega = - \beta \log \mathbb{Z}$ as $\mathcal{F} = \Omega + \mu \langle n \rangle$
By plugging in the chemical potential, we obtain that
\begin{equation}
\mathcal{F} = \frac{\langle n \rangle}{\beta} \log \left(\frac{-1\sqrt{1+8 \langle n \rangle e^{2 \beta  \epsilon
   }}}{4 e^{\beta  \epsilon }} \right) -\frac{e^{-2 \beta  \epsilon } \left(4 \langle n \rangle e^{2 \beta  \epsilon
   }+\sqrt{8 \langle n \rangle e^{2 \beta  \epsilon }+1}-1\right)}{8 \beta }\, .
- \langle n \rangle \log \langle n \rangle n \end{equation}
For large $\langle n \rangle$, the fluctuations of particles diminish, so only the states with the average number of particles become relevant and we can set $\langle n \rangle = n$. Moreover, the first term becomes dominant, so
\begin{equation}
 F(\beta,\epsilon,n) \approx \mu = \frac{\langle n \rangle}{\beta} \log \left(\frac{-1+\sqrt{1+8 \langle n \rangle e^{2 \beta  \epsilon
   }}}{4 e^{\beta  \epsilon }} \right) \, ,
\end{equation}
and we see that the free-energies of both approaches coincide for $c=n/2$.

\subsection{Derivation of the second law of thermodynamics for non-equilibrium structure-forming systems}

The time derivative of entropy can be expressed as
\begin{equation}
\frac{\mathrm{d} \mathcal{S}}{\mathrm{d}t} = - \sum_{ij} \dot{\wp}_{i}^{(j)} (\log \wp_{i}^{(j)}-1) - \sum_{ij} \dot{\wp}_i^{(j)}- \sum_{ij} \dot{\wp}_{i}^{(j)} \log \left(\frac{j!}{c^{j-1}}\right)\, .
\end{equation}
By plugging in the master equation we can further obtain that
{
\begin{eqnarray}
\dot{\mathcal{S}} &=& - \sum_{ijkl} w_{ik}^{jl} \wp_{k}^{(l)} \log \wp_{i}^{(j)} - \sum_{ijkl} w_{ik}^{jl} \wp_{k}^{(l)} \log \left(\frac{j!}{c^{j-1}}\right)\nonumber\\
&=&+\frac{1}{2}\sum_{ijkl}  (w_{ki}^{lj} \wp_{i}^{(j)} - w_{ik}^{jl} \wp_{k}^{(l)}) \log \frac{\wp_i^{(j)}}{\wp_k^{(l)}} + \frac{1}{2} \sum_{ijkl}  (w_{ki}^{lj} \wp_{i}^{(j)} - w_{ik}^{jl} \wp_{k}^{(l)}) \log \left(\frac{j!}{l!} c^{l-j} \right)\nonumber\\
&=&\underbrace{\frac{1}{2} \sum_{ijkl}  (w_{ki}^{lj} \wp_{i}^{(j)} - w_{ik}^{jl} \wp_{k}^{(l)}) \log \frac{w_{ki}^{lj} \wp_i^{(j)}}{w_{ik}^{jl} \wp_k^{(l)}}}_{\dot{\mathcal{S}}_i \geq 0} + \frac{1}{2} \sum_{ijkl}  (w_{ki}^{lj} \wp_{i}^{(j)} - w_{ik}^{jl} \wp_{k}^{(l)}) \log \left(\frac{j!}{l!} c^{l-j}  \frac{w_{ki}^{lj}}{w_{ik}^{jl}}\right)\nonumber\\
&=&\dot{\mathcal{S}}_i + \underbrace{\frac{\beta}{2}  \sum_{ijkl}  (w_{ki}^{lj} \wp_{i}^{(j)} - w_{ik}^{jl} \wp_{k}^{(l)})  (\epsilon_{i }^{(j)}-\epsilon_{k}^{(l)})}_{\dot{\mathcal{S}}_e =  \beta \dot{\mathcal{Q}}} + \frac{\alpha}{2} \sum_{ijkl}  (w_{ki}^{lj} \wp_{i}^{(j)} - w_{ik}^{jl} \wp_{k}^{(l)})(j-l)\, .
\end{eqnarray}
}%
Let us note that from the first law of thermodynamics,
\begin{equation}
\dot{\mathcal{U}} = \sum_{ij} \dot{\wp}_i^{(j)}\epsilon_{i}^{(j)} +  \sum_{ij} \wp_i^{(j)}\dot{\epsilon}_{i}^{(j)} = \dot{\mathcal{Q}} + \dot{\mathcal{W}} \, ,
\end{equation}
the entropy flow is equal to the heat flow over the temperature. Let us focus on last term, which can be expressed as
\begin{equation}
\frac{1}{2}\sum_{ijkl}  (w_{ki}^{lj} \wp_{i}^{(j)} - w_{ik}^{jl} \wp_{k}^{(l)})(l-j) = \sum_{ij} \dot{\wp}_i^{(j)} j { \equiv 0}
\end{equation}
is the time derivative of the normalization condition, i.e., the number of particles in the system { and therefore it is identical to zero.}
Therefore, the second law of thermodynamics can be expressed in the form
\begin{equation}\label{eq:2nd}
\frac{\mathrm{d} \mathcal{S}}{ \mathrm{d} t}
 =  \dot{\mathcal{S}}_i + \beta \dot{\mathcal{Q}} \, .
 \end{equation}
\subsection{Derivation of the detailed fluctuation theorem for non-equilibrium structure-forming systems}
{
Let us now focus on the derivation of entropy production along a stochastic trajectory.  We define entropy along a stochastic trajectory $|bt = (i(\tau),j(\tau))$ in the following form
\begin{equation}
s(\bt) = -  \left(\log \wp_{i(\tau)}^{(j(\tau))}(\tau) - 1 + \log \frac{(j(\tau))!}{c^{j(\tau)-1}}\right)
\end{equation}
Let us consider that the stochastic trajectory has jumps at times $t_z$ from $(i^-_z,j^-_z)$ to $(i^+_z,j^+_z)$ with transition rate $w_{i^+_z,i^-_z}^{j^+_z,j^-_z}$.
By taking the time derivative we obtain
\begin{eqnarray}
\dot{s} &=& - \frac{1}{\wp_{i(\tau)}^{(j(\tau))}} \partial_\tau \wp_{i(\tau)}^{(j(\tau))} -\sum_z \delta(\tau-t_z)  \log \left(\frac{\wp_{i_z^+}^{(j^+_z)}}{\wp_{i_z^-}^{(j^-_z)}}\right) - \sum_z \delta(\tau-t_z) \log \left(\frac{(j^+_z)!}{(j^-_z)!}\frac{c^{j^-_z-1}}{c^{j^+_z-1}}\right)\, .
\end{eqnarray}
With a little bit of care, we can recognize that the entropy can be decomposed into two terms. First is entropy flow rate
\begin{eqnarray}
\dot{s}_e &=& \sum_z \delta(\tau-t_z) \log \left(\frac{w_{i_z^- i_z^+}^{j_z^- j^+_z}}{w_{i_z^+ i_z^-}^{j_z^+ j^-_z}} \frac{(j^+_z)!}{(j^-_z)!}\frac{c^{j^-_z-1}}{c^{j^+_z-1}} \right) = \beta \sum_z \delta(\tau-t_z) (\epsilon_{i^-_z}^{(j^-_z)} - \epsilon_{i^+_z}^{(j^+_z)})
\end{eqnarray}
and second, entropy production rate along a stochastic trajectory
\begin{eqnarray}
\dot{s}_i &=&  - \frac{1}{\wp_{i(\tau)}^{(j(\tau))}} \partial_\tau \wp_{i(\tau)}^{(j(\tau))} - \sum_z \delta(\tau-t_z) \log \left(\frac{w_{i_z^- i_z^+}^{j_z^- j^+_z}\wp_{i_z^+}^{j_z^+}}{w_{i_z^+ i_z^-}^{j_z^+ j^-_z} \wp_{i_z^-}^{j_z^-}}\right)\, .
\end{eqnarray}
Thus, we obtain $\dot{s} = \dot{s}_i + \dot{s}_e$. The ensemble second law of thermodynamics can be recovered by multiplying the trajectory second law by {\janfin $\wp_i^{(j)}$ and summing over $i,j$.}}

Let us consider a stochastic trajectory $\bt = (i(\tau),j(\tau))$, where $\tau \in [0,T]$. Let us consider that jumps happen at times $\tau_z$, $z \in \{1,\dots,N\}$ from $(i_{z-1},j_{z-1})$ to $(i_{z},j_{z})$. Let us also time-dependent protocol $l(\tau)$ that controls the energy spectrum of the system. We define a quantity
\begin{equation}
\wpp(\bt) = \wp_{i_0}^{(j_0)}(0) \left[\prod_{z=1}^N e^{\int_{\tau_{z-1}}^{\tau_z}\ud {\tau'} w_{i_{z-1}i_{z-1}}^{j_{z-1}j_{z-1}}(l(\tau'))} w_{i_z i_{z-1}}^{{j_z j_{z-1}}}(l(\tau_z))\right]e^{\int_{\tau_N}^T \ud \tau' w_{i_N i_N}^{j_N j_N}(l(\tau'))}
\end{equation}
that corresponds to a ``probability'' of the trajectory $\bt$. {\janfin We can interpret this quantity as the relative number of clusters in the state $\bt$ over the total number of particles.} Indeed, this quantity does not sum up to one. Similarly, consider reverse trajectory $\btt = (i(T-\tau),j(T-\tau))$ and reverse protocol $\tilde{l}(\tau) = l(T-\tau)$. {\janfin We consider microreversibility, i.e., that the detailed balance holds also under the external protocol.} Then, we define \begin{equation}
\tilde{\wpp}(\btt) = e^{\int_{T-\tau_1}^T \ud \tau' w_{i_0i_0}^{j_0j_0}(\tilde{l}(\tau'))} \left[\prod_{z=1}^N w^{j_{z-1}j_z}_{i_{z-1}i_z}(\tilde{l}(T-\tau_z)) e^{\int_{T-\tau_{z+1}}^{T-\tau_z} \ud \tau' w^{j_z j_z}_{i_z i_z}(l(\tau'))}\right] \wp_{i_N}^{(j_N)}(T)\, .
\end{equation}
The log-ratio of both quantities can be expressed as
\begin{eqnarray}
\log \frac{\wpp(\bt)}{\tilde{\wpp}(\btt)} &=& \log \wp_{i_0}^{(j_0)}(0) - \log \wp_{i_N}^{(j_N)}(T) + \sum_{z=1}^N \log \frac{ w_{i_z i_{z-1}}^{{j_z j_{z-1}}}(l(\tau_z))}{w^{j_{z-1}j_z}_{i_{z-1}i_z}(\tilde{l}(T-\tau_z))}\nonumber\\
&=& \log \wp_{i_0}^{(j_0)}(0) + \log \frac{j_0!}{c^{j_0-1}} -  \log \wp_{i_N}^{(j_N)}(T) - \log \frac{j_N!}{c^{j_N-1}}\nonumber\\
&&+ \sum_{z=1}^N \log \left(\frac{w_{i_z i_{z-1}}^{{j_z j_{z-1}}}(l(\tau_z))}{w^{j_{z-1}j_z}_{i_{z-1}i_z}(\tilde{l}(T-\tau_z))} \frac{j_{z-1}!}{j_{z}!}\frac{c^{j_z-1}}{c^{j_{z-1}-1}} \right) = \Delta s - \Delta s_e = \Delta s_i
\end{eqnarray}

Let us now consider the master equation for probability $p_i^{(j)} = j \wp_i^{(j)}$ that a particle belongs to the cluster of size $j$ with energy $\epsilon_i^{(j)}$:
\begin{equation}
\dot{p}_i^{(j)} = \sum_{kl} W_{ik}^{jl} p_k^{(l)} \equiv \sum_{kl} \frac{j}{l}\,  w_{ik}^{jl} p_k^{(l)}
\end{equation}
From this, we obtain that $W_{ik}^{jl} = \frac{j}{l}\,  w_{ik}^{jl}$. Thus, the probability of observing the trajectory $\bt$ with the protocol $l(\tau)$ can be expressed as
\begin{equation}
\mathcal{P}(\bt) = p_{i_0}^{(j_0)}(0) \left[\prod_{z=1}^N e^{\int_{\tau_{z-1}}^{\tau_z}\ud {\tau'} W_{i_{z-1}i_{z-1}}^{j_{z-1}j_{z-1}}(l(\tau'))} W_{i_z i_{z-1}}^{{j_z j_{z-1}}}(l(\tau_z))\right]e^{\int_{\tau_N}^T \ud \tau' W_{i_N i_N}^{j_N j_N}(l(\tau'))}
\end{equation}
and similarly the probability of the time-reversed trajectory under the time-reversed protocol can be written as
\begin{equation}
\tilde{\mathcal{P}}(\btt) = e^{\int_{T-\tau_1}^T \ud \tau' W_{i_0i_0}^{j_0j_0}(\tilde{l}(\tau'))} \left[\prod_{z=1}^N W^{j_{z-1}j_z}_{i_{z-1}i_z}(\tilde{l}(T-\tau_z)) e^{\int_{T-\tau_{z+1}}^{T-\tau_z} \ud \tau' W^{j_z j_z}_{i_z i_z}(l(\tau'))}\right] p_{i_N}^{(j_N)}(T)\, .
\end{equation}
Therefore, the log-ratio can be expressed as
\begin{eqnarray}
\log \frac{\mathcal{P}(\bt)}{\tilde{\mathcal{P}}(\btt)} &=& \log p_{i_0}^{(j_0)}(0) - \log p_{i_N}^{(j_N)}(T) + \sum_{z=1}^N \log \frac{ W_{i_z i_{z-1}}^{{j_z j_{z-1}}}(l(\tau_z))}{W^{j_{z-1}j_z}_{i_{z-1}i_z}(\tilde{l}(T-\tau_z))}\nonumber\\
&=& \log p_{i_0}^{(j_0)}(0) - \log p_{i_N}^{(j_N)}(T) + \sum_{z=1}^N \log \left(\frac{j_z}{j_{z-1}}\right)^2 \frac{ w_{i_z i_{z-1}}^{{j_z j_{z-1}}}(l(\tau_z))}{w^{j_{z-1}j_z}_{i_{z-1}i_z}(\tilde{l}(T-\tau_z))}\nonumber\\
&=& \log p_{i_0}^{(j_0)}(0) - \log p_{i_N}^{(j_N)}(T) + 2 \log j_0 - 2 \log j_N + \sum_{z=1}^N \log \frac{ w_{i_z i_{z-1}}^{{j_z j_{z-1}}}(l(\tau_z))}{w^{j_{z-1}j_z}_{i_{z-1}i_z}(\tilde{l}(T-\tau_z))} \nonumber\\
&=& \log \frac{p_{i_0}^{(j_0)}(0)}{j_0} + \log \frac{j_0!}{c^{j_0}}  - \log \frac{p_{i_N}^{(j_N)}(T)}{j_N}  - \log \frac{j_N!}{c^{j_N}}
\nonumber\\&& + \log \frac{j_0}{j_N} + \sum_{z=1}^N \log \left(\frac{w_{i_z i_{z-1}}^{{j_z j_{z-1}}}(l(\tau_z))}{w^{j_{z-1}j_z}_{i_{z-1}i_z}(\tilde{l}(T-\tau_z))} \frac{j_{z-1}!}{j_{z}!}\frac{c^{j_z}}{c_{j_{z-1}}} \right)\nonumber\\
&=& \log \frac{\wpp(\bt)}{\tilde{\wpp}(\btt)} + \log \frac{j_0}{j_N}= \Delta s_i + \log \frac{j_0}{j_N}\, .
\end{eqnarray}
Therefore, we can write down that
\begin{equation}\label{eq:ftt}
\log \frac{\mathcal{P}(\bt)}{\tilde{\mathcal{P}}(\btt)} = \Delta s_i + \log \frac{j_0}{j_N}  = \Delta \sigma\, .
\end{equation}
We can express the probability of observing $\Delta \sigma$ as
\begin{eqnarray}
P(\Delta \sigma) &=& \int \mathcal{D}[\bt] \mathcal{P}(\bt) \delta \left(\Delta \sigma - \log \frac{\mathcal{P}(\bt)}{\tilde{P}(\btt)}\right) \nonumber\\
&=& \exp(\Delta {\janfin \sigma}) \int \mathcal{D}[\btt] \tilde{P}\left(\btt\right) \delta \left(-\Delta \sigma - \log \frac{\tilde{\mathcal{P}}\left(\btt\right)}{\tilde{P}(\bt)} \right)\nonumber\\
&=& \exp(\Delta \sigma) \tilde{P}(-\Delta \sigma)\, .
\end{eqnarray}
This gives us the detailed fluctuation theorem for $\Delta \sigma$.

Let us now assume that the initial state is in equilibrium. We rewrite $\Delta s_i$ as
\begin{equation}
\Delta s_i = \Delta s - \beta q = \beta(w+q-\Delta f) - \beta q + \Delta m = \beta w - \beta \Delta f\, .
\end{equation}
Let us now express free-energy $f$ of the equilibrium distribution
\begin{eqnarray}\label{eq:free}
f(\hat{\wp_i}^{(j)}) &=& \epsilon_{i}^{(j)} - T s(\wp_i^{(j)})  = \epsilon_{i}^{(j)} + T\left[\left(\log \frac{c^{j-1}}{j!} - j \alpha - \beta \epsilon_{i}^{(j)}\right) - 1 + \log\frac{j!}{c^{j-1}}\right]= - j \frac{\alpha}{\beta} - \frac{1}{\beta}\, .
\end{eqnarray}
From the ensemble averaging we obtain that
\begin{equation}
\sum_{ij} \hat{\wp}_i^{(j)} f(\hat{\wp}_i^{(j)}) = - \frac{\alpha}{\beta}  \sum_{ij} j \wp_i^{(j)} - \frac{1}{\beta} \sum_{ij}  \hat{\wp}_i^{(j)} =   - \frac{\alpha}{\beta} - \frac{1}{\beta} \sum_{ij} \hat{\wp}_i^{(j)} = - \frac{\alpha}{\beta} - \frac{\mathcal{M}}{\beta} = \mathcal{F}\, .
\end{equation}
By plugging \eqref{eq:free} into {\janfin \eqref{eq:ftt}} we obtain
\begin{equation}
\frac{\mathcal{P}(\bt)}{\tilde{\mathcal{P}}(\btt)} = \frac{j_0}{j_N} \exp\left(\beta w + (j_f \alpha_f - j_0 \alpha_0)  \right)\, .
\end{equation}
By substitution $j \exp(- \alpha j) = p^{(j)}/\mathcal{Z}_j$, we obtain
\begin{equation}
\frac{\mathcal{P}(\bt)}{\tilde{\mathcal{P}}(\btt)}  = \frac{p^{(j_0)}}{p^{(j_f)}} \frac{\mathcal{Z}_{j_f}}{\mathcal{Z}_{j_0}} \exp(\beta w)
\end{equation}
which can be further rewritten as
\begin{equation}
\frac{\mathcal{P}(\bt|j_0)}{\tilde{\mathcal{P}}(\btt|j_f)} = \exp(\beta w - \beta (\Phi_{j_f}({\janfin l}(T)) - \Phi_{j_0}({\janfin l}(0))
\end{equation}
where $\Phi_{j} = - \frac{1}{\beta} \log \mathcal{Z}_j$. The probability observing work $w$ starting from an equilibrium state with $j_0$ can be expressed as
\begin{eqnarray}
P_0(w|j_0) &=& \int \mathcal{D}[\bt] \mathcal{P}(\bt|j_0) \delta\left(\beta w - \beta (\Phi_{j_f} - \Phi_{j_0}) - \log \frac{\mathcal{P}(\bt|j_0)}{\tilde{\mathcal{P}}(\btt|j_f)}\right)\nonumber\\
&=& \exp(\beta w - \beta (\Phi_{j_f} - \Phi_{j_0})) \int \mathcal{D}[\btto] \mathcal{P}(\btt|j_f) \delta\left(-\beta w - \beta (\Phi_{j_0} - \Phi_{j_f}) - \log \frac{\tilde{\mathcal{P}}(\btt|j_f)}{\mathcal{P}(\bt|j_0)}\right)\nonumber\\
&=&  \exp(\beta w - \beta (\Phi_{j_f} - \Phi_{j_0})) \tilde{P}_0(-w|j_f)\,
\end{eqnarray}
which gives us the Crooks' work fluctuation theorem for structure-forming systems.
}

\subsection{Derivation of the self-consistency equation for magnetization in the fully connected Ising model}
The free-energy for the fully connected Ising model is given by
\begin{equation}
\mathcal{F} = - \frac{\alpha}{\beta}-\frac{\mathcal{M}}{\beta} \, ,
\end{equation}
where $\alpha$ are the same as for the molecule gas in the magnetic field (see the following section on the molecule gas in the presence of the magnetic field), just with the effective field $h_{eff} = (Jm+h)$. The self-consistency equation is obtained from the relation
\begin{equation}
m = -\frac{\partial \mathcal{F}}{\partial h}|_{h=0} \, ,
\end{equation}
which leads to the following equation
\begin{eqnarray}\label{eq:sc}
m = \frac{\sinh(J m \beta)}{\sqrt{n+\cosh(J m \beta)^2}} \left(1+\frac{n}{\left(\cosh (\beta  J m)+\sqrt{\cosh ^2(\beta  J m)+n}\right)^2}\right)\, .
\end{eqnarray}
This equation has to be solved numerically, similarly to the case of the
fully connected Ising model without molecule states.
The solution is depicted in Fig. 4 in the main text.

\subsection{Monte Carlo simulation of the fully connected Ising model}
We describe now Monte Carlo simulation applied to a system of free particles with two states $\{\uparrow,\downarrow\}$ and the two-particle molecule with one state $\{\|\}$ and a Hamiltonian $H(n_\uparrow,n_\downarrow,n_{\|}) = -h(n_\uparrow - n_\downarrow)$. The algorithm puts particles into two boxes, one box for free particles and one box for two particle molecules. The approach is similar to how Panagiotopoulos described in \cite{Panagiotopoulos99}.
For example consider boxes, one for ``Atoms'' and one for two particle molecules. Here Box 1 contains the states $\{\uparrow,\downarrow\}$ and Box 2 the states $\{\|\}$.
Two kinds of MC-moves are tried. The first kind of move randomly chooses a particle in the
$\uparrow$ or $\downarrow$ state and changes it to the other state.
The move is then accepted with the probability $\min\left(1,\exp(-\beta \Delta H) \right)$.
The second kind of move is either dividing a state $\{\|\}$ particle in two states $\{\uparrow,\downarrow\}$,
or combining two random particles from Box 1 to a state $\{\|\}$ particle.
Which one of the two moves is tried is chosen randomly with equal probability.
The division move takes a Box 2 particle (two particle molecule) and deletes it and two new Box 1 particle (one-atomic particles) are generated in its stead.
The state of the new particle is either $\uparrow$ or ${\downarrow}$.
Which of the two states the particles are created in is chosen randomly
with the probability of the current distribution of the two states in Box 1.
The move is then accepted with the probability $\min\left(1,\frac{2 n_{\|} n_{\uparrow}! n_{\downarrow}!}{\tilde{n}_{\uparrow}! \tilde{n}_{\downarrow}!} \exp(-\beta \Delta H) \right)$,
where $\tilde{n}_{\uparrow}$ and $\tilde{n}_{\downarrow}$
are the numbers for  $\uparrow$ and ${\downarrow}$ particles after the move, respectively.
The combination move takes two random particles in Box 1 deletes them and creates a new particle in Box 2.
The  move is then accepted with  $\min\left(1,\frac{n_{\uparrow}! n_{\downarrow}!}{\tilde{n}_{\uparrow}! \tilde{n}_{\downarrow}! (2 n_{\|}+2)} \exp(-\beta \Delta H) \right)$.
Here, $\tilde{n}_{\uparrow}$ and $\tilde{n}_{\downarrow}$ are once again the number of  $\uparrow$ and ${\downarrow}$ particles  after the move, respectively. The simulation ends after the ensemble does not change significantly
anymore and reaches an equilibrium.
The final output is the mean over 1000 simulations at a certain temperature value. Each simulation consists of 2 million steps.

\section{Supplementary discussion}

\subsection{Relation of the entropy for molecule system to axiomatic frameworks}
In this section, we discuss the relation of the entropy for molecule systems to existing axiomatic frameworks, including axiomatics of Shannon and Khinchin and its generalizations according to Tempesta and Jensen, and Hanel and Thurner, and also axiomatics according to Shore and Johnson.

\paragraph*{Lieb-Yngvason axioms}
Let's discuss the main properties of the entropic functional \eqref{eq:ent}. We start with \emph{additivity} and \emph{extensivity}, as  introduced by Lieb and Yngvason
\cite{Lieb}.  \emph{Additivity} can be formulated as $S((X,Y)) = S(X)+S(Y)$ (see Eq. (2.4) in \cite{Lieb}) where $(X,Y)$ is a cartesian product of two systems, i.e., a state of the composed systems is a
{ pair}
$(x,y)$, where $x \in X$ and $y \in Y$. We consider two systems, one with $\chi$ particles, the other with $\xi$ particles. For simplicity, consider that the particles can attain the same states for both systems.
We denote the number of clusters in state $x_i^{(j)}$ in the first {and second subsystem as $\chi_i^{(j)}$ and} $\xi_i^{(j)}$, respectively. Since the two subsystems are independent, the total multiplicity for $n_i^{(j)} =\chi_i^{(j)}+\xi_i^{(j)}$ is simply given by the product of two multiplicities, so that
\begin{equation}
W(n_i^{(j)}) = W(\chi_i^{(j)}) W(\xi_i^{(j)}) \, ,
\end{equation}
from which we immediately see that $S(n_i^{(j)}) = S(\chi_i^{(j)})+S(\xi_i^{(j)})$.
The second property, \emph{extensivity}, states that $S(tX) = t S(X)$, where $tX$ is the rescaled version of the system (see Eq. (2.5) in \cite{Lieb}). For simplicity, let us consider that we double the system, i.e., $t=2$. This means that we have $2n$ particles and there are $2 n_{i}^{(j)}$ particles in the state $x_i^{(j)}$.  We also have to double the systems' volume, i.e., we divide the total system into $2b$ boxes. The entropy of the double system is then equal to
\begin{eqnarray}
S(2n) &=& 2n \log \frac{2n}{2b} - 2n - \sum_{ij} 2 n_i^{(j)} \left(\log \frac{2 n_i^{(j)}}{2b}-1\right)\nonumber\\ &&- \sum_{ij} 2 n_i^{(j)} \log j! = 2 S(n) \, .
\end{eqnarray}
Thus, the entropy is extensive. Another important property is \emph{concavity} of entropy, ensuring the uniqueness of the maximum entropy principle. Since it is straightforward to show that
\begin{equation}
\frac{\partial^2 S(P)}{\partial p_{i}^{(j)} \partial p_{i'}^{(j')}} = -\frac{1}{j p_i^{(j)}} \delta_{ii'} \delta_{jj'} \, ,
\end{equation}
we conclude that the entropy is a \emph{concave} function of probability distribution.

\paragraph*{Classes of entropies and joint entropies:}
Before we move to particular axiomatic frameworks based on information-theoretic approaches, let us discuss typical classes of entropies that are taken into account. These are
\begin{enumerate}
\item \emph{Trace-class entropies} $S(P) = \sum_i g(p_i)$
\item \emph{Sum-class entropies} $S(P) = f(\sum_i g(p_i))$.
\end{enumerate}
These entropies are widely used in information theory and statistical physics because of its nice properties. However, neither of the classes is suitable for our case. One of the issues that occur is that these entropy classes are symmetric functions of all probabilities, which is a consequence of the principle that relabeling the states should not change the entropy (also called permutational invariance - see the next section about Shore-Johnson axioms). It is, however, not the case of a system with molecule states. Here, switching the order of molecules (e.g., from free particle states to molecule states) changes the system's entropy since the states corresponding to molecules of different orders are states of different types, and one cannot expect that the symmetry argument holds.

There is another assumption that is implicitly considered by these classes of entropy. Namely, it is the assumption that the \emph{joint entropy}, i.e., the entropy of the joint distribution has the same functional form as the entropy of the marginal distribution, i.e., in the case of the joint probability of two random variables, we have
\begin{equation}
S(p_{ij}) = f(\sum_{ij} g(p_{ij}))
\end{equation}
This assumption makes perfect sense for systems with exponential sample spaces, where the joint distribution can also be interpreted as a marginal distribution of a system that is obtained by merging the two systems together, i.e., $X \times X \sim 2X$, where on the left-hand side is the cartesian product of two systems and on the right-hand side is the rescaled version of the system. However, in our case $W(2n) \geq W(n)^2$ so there is no such correspondence.

Let us demonstrate this on an example of two systems $A$ and $B$ with $n$ and $m$ molecules. Corresponding probability distributions are $\tilde{u}_i^{(j)}$ and $\tilde{v}_{i'}^{(j')}$. The entropy of the composed system $(A,B) \equiv A \times B$ is given by the sum of the entropies of two systems, as demonstrated above. If we try to express this entropy in terms of joint distribution $\tilde{p}_{ii'}^{(jj')} = \tilde{u}_i^{(j)} \tilde{v}_{i'}^{(j')}$ we obtain

\begin{eqnarray}
S(A,B) &=&  - \sum_{ij} \frac{\tilde{u}_i^{(j)}}{j} \left(\log \frac{\tilde{u}_i^{(j)}}{j}-1\right) - \sum_{ij} \frac{\tilde{u}_i^{(j)}}{j} \log\left(\frac{j!}{c_A^{j-1}}\right)\nonumber\\
&&- \sum_{i'j'} \frac{\tilde{v}_{i'}^{(j')}}{j'} \left(\log \frac{\tilde{v}_{i'}^{(j')}}{j'}-1\right) - \sum_{i'j'} \frac{\tilde{v}_{i'}^{(j')}}{j'} \log\left(\frac{j'!}{c_B^{j'-1}}\right)\nonumber\\
&=& - \sum_{iji'j'} \tilde{p}_{ii'}^{(jj')} \left(\frac{1}{j} \log \frac{\sum_{i'j'} \tilde{p}_{ii'}^{(jj')}}{j}-1+\frac{1}{j'} \log \frac{\sum_{ij} \tilde{p}_{ii'}^{(jj')}}{j'}-1\right)\nonumber\\
&& - \sum_{iji'j'} \tilde{p}_{ii'}^{(jj')} \left(\frac{1}{j} \log \frac{j!}{c_A^{j-1}} + \frac{1}{j'} \log \frac{j'!}{c_B^{j'-1}}\right)
\end{eqnarray}
So we see that the joint entropy is expressible in terms of the joint distribution, but the functional form is different from the entropy of the marginal distribution.

Let us note that the composed system created from systems with $n_1$ particles and $n_2$ particles is, in general, different from a system with $n_1+n_2$ particles. Therefore, one cannot expect that the system's entropy can be obtained as a sum of the subsystems. The system's state space with $n_1+n_2$ particles cannot be represented as a cartesian product of states from the subsystems. A simple example can exemplify this issue: we can consider a molecule state consisting of two particles in a system. The first particle is taken from the first $n_1$ particles, and the other particle is taken from the remaining $n_2$ particles. Such a state has no representation in the Cartesian product. It is the consequence of the fact that the sample space grows super-exponentially, and therefore $W(n) > W(n_1)W(n_2)$.

\paragraph*{Shannon-Khinchin axioms:}
Shannon-Khinchin (SK) axioms characterize the properties of the Shannon entropy from the information-theoretic point of view. They were proposed independently by Shannon \cite{Shannon48} and Khinchin \cite{Khinchin57} to determine the Shannon entropy uniquely. In the original formulation, the Shannon-Khinchin axioms are the following:
\begin{enumerate}
  \item \emph{Continuity:} Entropy is a continuous function of probability distribution.
  \item \emph{Maximality:} Entropy is maximal for the uniform distribution.
  \item \emph{Expansibility:} Adding an elementary event with probability zero does not change the entropy.
  \item \emph{Additivity:} $H(A,B) = H(A)+H(B|A)$, where $H(B|A) = \sum_i p_i H(B|A=a_i)$.
\end{enumerate}
Since the original four axioms uniquely determine Shannon entropy $H(P) = - \sum_i p_i \log p_i$, several authors proposed generalizations of the original scheme. The typical approach is to weaken the fourth axiom to obtain a wider class of entropies while the first three axioms remain unchanged. Thus, before discussing the generalizations of the fourth SK axiom, let us focus on the first three axioms.

It is easy to show that the molecule entropy fulfills the first and the third axiom. As we have discussed in the main text, the entropy is not maximized by the uniform distribution when molecules of different sizes are present. The maximality axiom results from a similar requirement, i.e., that the entropic functional should be symmetric and Schur-concave function of the probability distribution. In the last section, we have discussed that the assumption of symmetry is not suitable for our system. Therefore, it is more natural to weaken the maximality axiom
to the following form:
\begin{enumerate}
\setcounter{enumi}{1}
\item \emph{Maximality:} Entropy is maximal for a distribution, where each microstate contributes with equal probability.
\end{enumerate}
By a generalization of SK2, the entropy of molecule states fulfills the first three SK axioms, plus the first part of the fourth one, i.e., $S(A \times B) = S(A) + S(B)$, where $A \times B$ is the Cartesian product of random variables. It means that the state space is a Cartesian product of $A$ and $B$ and joint probability is simply a marginal probabilities product, as demonstrated in the previous section, when we discussed entropy's additivity. Let us now omit discussion about the definition of conditional probability, which is quite technical and would deserve a separate paper.

\paragraph*{Tempesta axioms and group entropies:}
Tempesta proposed in the series of papers \cite{Tempesta16,Jensen18a,Tempesta20} a generalization of SK axioms, but weakening the fourth axiom by imposing that the entropy should fulfill the group property:
\begin{enumerate}
\setcounter{enumi}{3}
\item \emph{Group composability:} $S(A,B) = \Phi(S(A),S(B))$
\end{enumerate}
where $x,y \mapsto \Phi(x,y)$ is the group action. In the previous section, we have shown that the entropy of molecule systems is additive and therefore fulfills the composability property with $\Phi(x,y) = x+y$. However, the whole calculation was done in the framework of sum-class of entropies. Therefore, the entropy of structure-forming systems does not fall into this reduced class of entropies that work with states of the same structure. However, when relaxing the second SK axiom and omitting the requirement of symmetric entropies, the entropy of molecule systems belongs to the class of group entropies. Relaxing 2nd SK axiom in the framework of group entropies would an important step in the future research.

\paragraph*{Hanel-Thurner axioms and entropy scaling:}
Hanel and Thurner generalized the fourth axiom differently. They did not require any particular composition law; they only examined the asymptotic scaling of the entropy for the case of distribution that maximizes the entropy as a function of the system size (here the number of particles) \cite{Hanel11a,Korbel18}. The resulting classification leads to the set of scaling exponents (originally $(c,d)$) that determined universality classes of entropic functionals. Since the entropic functionals are considered to be trace-class, asymptotic scaling was examined for uniform distribution. Nevertheless, for the entropy of structure-forming systems, the distribution that maximizes the entropy does not have to be uniform. Actually, by plugging the MaxEnt distribution into the entropy of molecule systems, we obtain that $S(n) \sim \log n$, so we obtain that $(c,d) = (0,1)$, which is the case of additive entropies (including Shannon entropy).

\paragraph*{Shore-Johnson axioms:}
Shore and Johnson considered the principle of maximum entropy as a statistical inference method and formulated a set of consistency requirements \cite{Jizba19,Korbel20}. They considered the class of inductive inference (i.e., they are in the form of averaged quantities). The requirements are the following:
\begin{itemize}
\item \emph{Uniqueness:} the result should be unique.
\item \emph{Permutation invariance:} the permutation of states
should not matter.
\item \emph{Subset independence:} It should not matter whether one
treats disjoint subsets of system states in terms of separate
conditional distributions or in terms of the full distribution.
\item \emph{System independence:} It should not matter whether one
accounts for independent constraints related to independent
systems separately in terms of marginal distributions or in
terms of full system.
\item \emph{Maximality:} In absence of any prior information, the
uniform distribution should be the solution.
\end{itemize}
In \cite{Korbel20} was shown, these axioms are equivalent to Tempesta's group composability under the assumption of sum-class entropies. Indeed, some of the axioms are not fulfilled for the case of molecule entropy. Especially axioms 2 and 5 are other forms of the same assumption that entropy should be a symmetric function of its variables. Therefore, a generalization of SJ axioms for state spaces with states of different types should be reasonable.

\subsection{Presence of magnetic gas phase for molecule-forming particles in presence of magnetic field}
\begin{figure}[t]
\includegraphics[width=8cm]{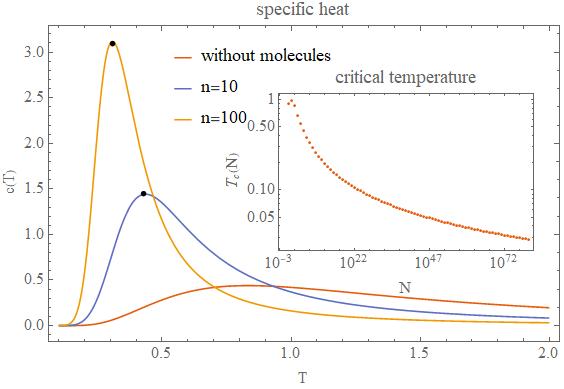}\\
\includegraphics[width=3cm]{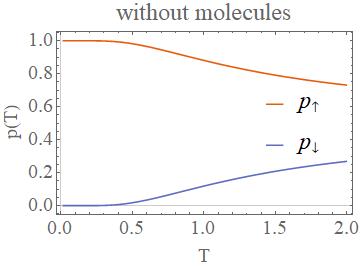}
\includegraphics[width=3cm]{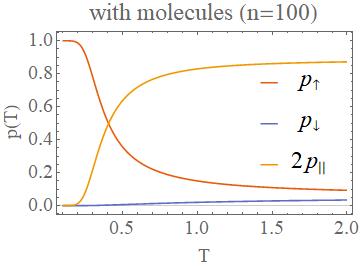}\\
\caption{Specific heat of the molecule model with magnetic field.
We observe a phase transition for the magnetic model. The critical temperature decreases with $n$ to zero very slowly (inset).}
\label{fig1}
\end{figure}
Let us consider a system of $n$ particles where free particles can have two states $\{\uparrow,\downarrow\}$
and two-particle molecules have one state $\{\|\}$. Let us assume the case when the system is small and dense so that all particles can interact with each other. The Hamiltonian corresponding to the magnetic field is
\begin{equation}
H(n_\uparrow,n_\downarrow,n_{\|}) = -h(n_\uparrow - n_\downarrow) \, .
\end{equation}
We calculate the specific heat $c = -T \frac{\mathrm{d}^2 F}{\mathrm{d} T^2}$  and we see that there is a phase transition between the magnetic phase and molecule phase which grows with $n$, as shown in Fig. \ref{fig1} Let us note that the dependence of critical temperature on $n$,
where $\lim_{N \rightarrow \infty} T_c(N) = 0$. However, as shown in the inset,
the critical temperature is well separated from zero even for large, but finite systems,
since the convergence is very slow. It should be mentioned that the magnetic gas has been observed for low temperatures
experimentally \cite{Jo09,Duine05}.

\subsection{Finite-size correction to chemical potential}
Let us consider again the chemical reaction $2 X \rightleftharpoons X_2$. Without loss of generality, assume
that free particles carry some energy $\epsilon$. The equilibrium constant of the chemical reaction can be expressed as
\begin{equation}
K_c = \frac{n_{X_2}}{n_X^2} = \frac{\wp_{X_2}}{(\wp_X)^2} = \exp\left(2 \beta \epsilon +  \log \frac{c}{2}\right) \, .
\end{equation}
Thus, we obtain the effective chemical potential, \mbox{$ \Delta \mu^{eff} = 2 \epsilon + \frac{1}{\beta} \log \frac{c}{2}$},
where the first term can be obtained from the ordinary grand-canonical ensemble of
two-gas system and the second one is the correction which is obtained from the molecule entropy.
This means that if the two gases are perfectly mixed in a small region so that every particle can interact
with each other particle, the value of the chemical potential explicitly depends on the number of particles ---
with an increasing number of particles the chemical potential increases.
In fact, the finite-size corrections to the chemical potential have been considered in several aspects,
especially in case of interacting particles \cite{Smit89,Siepman92}.
In our case, the correction is simply because of the structure-forming states.

\section{Supplementary References}


\begin{thebibliography}{99}
\bibitem{Thurner18}
S. Thurner, P. Klimek, and R. Hanel,
Introduction to the
Theory of Complex Systems.
Oxford University Press, Oxford (2018).

\bibitem{Hanel11a}
R. Hanel and S. Thurner,
A comprehensive classification of complex statistical systems and an axiomatic derivation of their entropy and distribution functions. \emph{Europhys. Lett.} \textbf{93} (2011) 20006.

\bibitem{Hanel11b}
R. Hanel and S. Thurner,
When do generalized entropies apply? How phase space volume determines entropy.
\emph{Europhys. Lett.} \textbf{96} (2011) 50003.

\bibitem{Hanel14}
R. Hanel, S. Thurner, and M. Gell-Mann,
How multiplicity determines entropy and the derivation of the maximum entropy principle for complex systems.
\emph{Proc. Natl. Acad. Sci. U.S.A.} \textbf{111}, 6905 (2014).

\bibitem{Korbel18}
J. Korbel, R. Hanel and S. Thurner,
Classification of complex systems by their sample-space scaling exponents.
\emph{New J. Phys.} \textbf{20} (2018) 093007.

\bibitem{Tsallis88}
C. Tsallis,
Possible generalization of Boltzmann-Gibbs statistics.
\emph{J. Stat. Phys.} \textbf{52} (1988) 479.

\bibitem{Rajagopal96}
A.K. Rajagopal,
Dynamic linear response theory for a nonextensive system based on the Tsallis prescription.
\emph{Phys. Rev. Lett.} \textbf{76} (1996) 3469.

\bibitem{Kaniadakis02}
G. Kaniadakis,
Statistical mechanics in the context of special relativity.
\emph{Phys. Rev. E} \textbf{66} (2002) 056125.

\bibitem{Jizba04}
P. Jizba and T. Arimitsu,
The world according to Renyi: thermodynamics of multifractal systems.
\emph{Ann. Phys.} \textbf{312} (2004) 17.

\bibitem{Anteneodo04}
C. Anteneodo and A. R. Plastino,
Maximum entropy approach to stretched exponential probability distributions.
\emph{J. Phys. A} \textbf{32(7)} (1999) 1089.

\bibitem{Lutz13}
E. Lutz and F. Renzoni,
Beyond Boltzmann-Gibbs statistical mechanics in optical lattices.
\emph{Nat. Phys.} \textbf{9} (2013) 615-619.

\bibitem{Dechant15}
A. Dechant, D.A. Kessler and E. Barkai,
Deviations from Boltzmann-Gibbs statistics in confined optical
lattices.
\emph{Phys. Rev. Lett.} \textbf{115} (2015) 173006.

\bibitem{Jizba19}
P. Jizba and J. Korbel,
Maximum Entropy Principle in Statistical Inference: Case for Non-Shannonian Entropies.
\emph{Phys. Rev. Lett.} \textbf{122} (2019) 120601.

\bibitem{Korbel20}
P. Jizba and J. Korbel,
When Shannon and Khinchin meet Shore and Johnson:
Equivalence of information theory and statistical inference axiomatics.
\emph{Phys. Rev. E} \textbf{101} (2020) 042126.

\bibitem{Jensen18}
H. J. Jensen, R. H. Pazuki, G. Pruessner and P. Tempesta,
Statistical mechanics of exploding phase spaces: ontic open systems.
\emph{J. Phys. A} \textbf{51} (2018) 375002.

\bibitem{Latora17}
V. Latora, V. Nicosia, and G. Russo,
Complex networks: principles, methods and applications. Cambridge University Press, Cambridge, (2017).

\bibitem{Squartini15}
T. Squartini, J. de Mol, F. den Hollander and D. Garlaschelli,
Breaking of Ensemble Equivalence in Networks.
\emph{Phys. Rev. Lett.} \textbf{115} (2015) 268701.

\bibitem{Berge73}
C. Berge,
Graphs and Hypergraphs.
North-Holland mathematical library (1973).

\bibitem{Temkin96}
O. N. Temkin, A. V. Zeigarnik, and D. G. Bonchev,
Chemical reaction networks: a graph-theoretical approach.
CRC Press (1996).

\bibitem{Flamm15}
C. Flamm, B. M. R. Stadler, and P. F. Stadler, Generalized topologies: hypergraphs, chemical reactions, and biological evolution.
Advances in Mathematical Chemistry and Applications. Bentham Science Publishers, (2015) 300-328.

\bibitem{Smit89}
B. Smit and D. Frenkel,
Explicit expression for finite size corrections to the chemical potential.
\emph{J. Phys.: Condens. Matt.} \textbf{1} (1989) 8659.

\bibitem{Siepman92}
J. I. Siepmann, I. R. McDonald, and D. Frenkel,
Finite-size corrections to the chemical potential.
\emph{J. Phys.: Condens. Matt.} \textbf{4(3)} (1992) 679.

\bibitem{Chandler76}
D. Chandler and L. R. Pratt,
Statistical mechanics of chemical equilibria and intramolecular structures of nonrigid molecules in condensed phases.
\emph{J. Chem. Phys.} \textbf{65(8)} (1976) 2925-2940.

\bibitem{Kreuzer81}
H. J. Kreuzer,
Nonequilibrium thermodynamics and its statistical foundations.
Clarendon Press, Oxford and New York (1981).

\bibitem{Cummings84}
P. T. Cummings and G. Stell,
Statistical mechanical models of chemical reactions: Analytic solution of models of $A+ B \rightleftharpoons AB$ in the Percus-Yevick approximation.
\emph{Mol. Phys.} \textbf{51(2)} (1984) 253-287.

\bibitem{Schmiedl07}
T. Schmiedl and U. Seifert,
Stochastic thermodynamics of chemical reaction networks.
\emph{J. Chem. Phys.} \textbf{126(4)} (2007) 044101.

\bibitem{Israelachvili77}
J. N. Israelachvili, D. J. Mitchell, and J. W. Ninham,
Theory of self-assembly of lipid bilayers and vesicles.
\emph{Biochimica et Biophysica Acta (BBA)-Biomembranes} {\bf 470(2)} (1977) 185-201.

\bibitem{Aranson06}
I. S. Aranson, and L. S. Tsimring,
Theory of self-assembly of microtubules and motors.
\emph{Phys. Rev. E} {\bf 74(3)} (2006) 031915.

\bibitem{Walther13}
A. Walther and A. H. E. Muller,
Janus particles: synthesis, self-assembly, physical properties, and applications.
\emph{Chem. Rev.} {\bf 113(7)} (2013) 5194-5261.

\bibitem{Grabow14}
W. W. Grabow and L. Jaeger,
RNA self-assembly and RNA nanotechnology.
\emph{Accounts of chemical research} {\bf 47(6)} (2014) 1871-1880.


\bibitem{Fantoni11}
R. Fantoni,  A. Giacometti, F. Sciortino  and  G. Pastore, Cluster theory of Janus particles. \emph{Soft Matter} {\bf 2011(7)} (2011) 2419-2427.


\bibitem{Nguyen16}
M. Nguyen and S. Vaikuntanathan,
Design principles for nonequilibrium self-assembly.
\emph{Proc. Nat. Acad. Sci. USA} {\bf 113 (50)} (2016) 14231-14236.

\bibitem{Bisker18}
G. Bisker and J. L. England,
Nonequilibrium associative retrieval of multiple stored self-assembly targets-
\emph{Proc. Nat. Acad. Sci. USA} {\bf 115 (45)} (2018) E10531-E10538.

\bibitem{Restrepo19}
A. Arango-Restrepo,  D. Barrag\'{a}n  and  J. M. Rubi,
Self-assembling outside equilibrium: emergence of structures mediated by dissipation.
\emph{Phys. Chem. Chem. Phys.} {\bf 21} (2019) 17475-17493.

\bibitem{Boltzmann}
L. Boltzmann,
\"{U}ber das Arbeitsquantum, welches bei chemischen Verbindungen gewonnen werden kann.
{\em Annalen der Physik} {\bf 258(5)} (1884) 39-72.




\bibitem{Lieb}
E. H. Lieb and J. Yngvason,
The physics and mathematics of the second law of thermodynamics.
\emph{Phys. Rep.} \textbf{310(1)} (1999) 1-96.

\bibitem{Likos16}
 C. N. Likos, F. Sciortino, E. Zaccarelli and P. Ziherl, Soft Matter Self-Assembly. \emph{Proceedings of the International School of Physics "Enrico Fermi"} {\bf 193} (2016).

\bibitem{Vissersa14}
T. Vissersa, F. Smallenburga, G. Munao, Z. Preisler and F. Sciortino, Cooperative polymerization of one-patch colloids.
\emph{J. Chem. Phys.} {\bf 140} (2014) 144902.


\bibitem{Preisler14}
Z. Preisler, T. Vissers, G. Munao, F. Smallenburg, and F.. Sciortino,
Equilibrium phases of one-patch colloids with short-range attractions.
\emph{Soft Matter} {\bf 10} (2014) 5121-5128.

\bibitem{Kern03}
N. Kern and D. Frenkel,
Fluid–fluid coexistence in colloidal systems with short-ranged strongly directional attraction,
\emph{J. Chem. Phys.} {\bf 118} (2003) 9882–9889.

\bibitem{Rovigatti18}
L. Rovigatti, J. Russo, and F. Romano,
How to simulate patchy particles.
\emph{Eur. Phys. J. E} {\bf 41} (2018) 59.

\bibitem{bimodality}
R. Pfister, K. Schwarz, M. Janczyk, R. Dale, and J. Freeman, Good things peak in pairs: a note on the bimodality coefficient.
\emph{Front. Psychol.} \textbf{4} (2013) 700.

\bibitem{Griffiths66}
R. B. Griffiths, C.-Y. Weng, and J. S. Langer,
Relaxation Times for Metastable States in the Mean-Field Model of a Ferromagnet.
\emph{Phys. Rev.} \textbf{149} (1966) 301.

\bibitem{Botet82}
 R. Botet, R. Jullien, and P. Pfeuty,
 Size Scaling for Infinitely Coordinated Systems.
\emph{Phys. Rev. Lett.} \textbf{49} (1982) 478.

\bibitem{Gulbahce04}
N. Gulbahce, H. Gould, and W. Klein,
Zeros of the partition function and pseudospinodals in long-range Ising models.
\emph{Phys. Rev. E} \textbf{69}  (2004) 036119.

\bibitem{Romano14}
L. Colonna-Romano, Harvey Gould, and W. Klein,
Anomalous mean-field behavior of the fully connected Ising model.
\emph{Phys. Rev. E} \textbf{90(4)} (2014) 042111.

\bibitem{Seifert08}
U. Seifert,
Stochastic thermodynamics: principles and perspectives.
\emph{Eur. Phys. J. B} \textbf{64} (2008) 423–431.

\bibitem{Esposito10}
M. Esposito and C. Van den Broeck,
The Three Faces of the Second Law: I. Master Equation Formulation.
\emph{Phys. Rev. E} \textbf{82} (2010) 011143.


\bibitem{Esposito10a}
M. Esposito and C. Van den Broeck,
Three Detailed Fluctuation Theorems.
\emph{Phys. Rev. Lett.} \textbf{104} (2010) 090601.


\bibitem{Crooks99}
G. E. Crooks,
Entropy production fluctuation theorem and the nonequilibrium work relation for free-energy differences.
\emph{Phys. Rev. E} \textbf{60(3)} (1999) 2721.



\bibitem{Kagan09}
D. Kagan et al.,
Chemical sensing based on catalytic nanomotors: motion-based detection of trace silver.
\emph{J. Am. Chem. Soc.} \textbf{131(34)} (2009) 12082-12083.



\end{thebibliography}

\begin{thebibliography}{99}

\bibitem{Panagiotopoulos99}
A. Z. Panagiotopoulos,
Monte Carlo methods for phase equilibria of fluids.
\emph{J. Phys.: Condens. Matt.} \textbf{12(3)} (1999) 25-52.


\bibitem{Lieb}
E. H. Lieb and J. Yngvason,
The physics and mathematics of the second law of thermodynamics.
\emph{Phys. Rep.} \textbf{310(1)} (1999) 1-96.


\bibitem{Shannon48} C. Shannon,
A mathematical theory of communication.
\emph{Bell Syst. Tech. J.} \textbf{27} (1948) 379.

\bibitem{Khinchin57}
A.I. Khinchin,
Mathematical Foundations of Information Theory.
Dover Publications, New York (1957).

\bibitem{Tempesta16}
P. Tempesta,
Beyond the Shannon–Khinchin formulation: The composability axiom and the universal-group entropy.
\emph{Ann. Phys.} \textbf{365} (2016) 180.

\bibitem{Jensen18a}
H. J. Jensen and P. Tempesta,
Group entropies: From Phase Space Geometry to Entropy Functionals via Group Theory.
\emph{Entropy} \textbf{20} (2018) 804.

\bibitem{Tempesta20}
P. Tempesta and H. J. Jensen,
Universality Classes and Information-Theoretic Mesure of Complexity via Group Entropies.
\emph{Sci. Rep.} \textbf{10} (2020) 5952.


\bibitem{Hanel11a}
R. Hanel and S. Thurner,
A comprehensive classification of complex statistical systems and an axiomatic derivation of their entropy and distribution functions. \emph{Europhys. Lett.} \textbf{93} (2011) 20006.

\bibitem{Korbel18}
J. Korbel, R. Hanel and S. Thurner,
Classification of complex systems by their sample-space scaling exponents.
\emph{New J. Phys.} \textbf{20} (2018) 093007.


\bibitem{Jizba19}
P. Jizba and J. Korbel,
Maximum Entropy Principle in Statistical Inference: Case for Non-Shannonian Entropies.
\emph{Phys. Rev. Lett.} \textbf{122} (2019) 120601.


\bibitem{Korbel20}
P. Jizba and J. Korbel,
When Shannon and Khinchin meet Shore and Johnson:
Equivalence of information theory and statistical inference axiomatics.
\emph{Phys. Rev. E} \textbf{101} (2020) 042126.

\bibitem{Jo09}
G.-B. Jo, Y.-R. Lee, J.-H. Choi, C. A. Christensen, T. H. Kim, J. H. Thywissen, D. E. Pritchard, and W. Ketterle,
Itinerant Ferromagnetism in a Fermi Gas of Ultracold Atoms. \emph{Science} \textbf{325} (2009)  5947.

\bibitem{Duine05}
R. A. Duine and A. H. MacDonald
Itinerant Ferromagnetism in an Ultracold Atom Fermi Gas.-
\emph{Phys. Rev. Lett.} \textbf{95} (2005) 230403.

\bibitem{Smit89}
B. Smit and D. Frenkel,
Explicit expression for finite size corrections to the chemical potential.
\emph{J. Phys.: Condens. Matt.} \textbf{1} (1989) 8659.

\bibitem{Siepman92}
J. I. Siepmann, I. R. McDonald, and D. Frenkel,
Finite-size corrections to the chemical potential.
\emph{J. Phys.: Condens. Matt.} \textbf{4(3)} (1992) 679.




\end{thebibliography}
\end{document}